\documentclass{book}
\newdimen\footheight
\makeatletter
\newcommand\vpt   {\edef\f@size{\@vpt}\rm}
\newcommand\vipt  {\edef\f@size{\@vipt}\rm}
\newcommand\viipt {\edef\f@size{\@viipt}\rm}
\newcommand\viiipt{\edef\f@size{\@viiipt}\rm}
\newcommand\ixpt  {\edef\f@size{\@ixpt}\rm}
\newcommand\xpt   {\edef\f@size{\@xpt}\rm}
\newcommand\xipt  {\edef\f@size{\@xipt}\rm}
\newcommand\xiipt {\edef\f@size{\@xiipt}\rm}
\newcommand\xivpt {\edef\f@size{\@xivpt}\rm}
\newcommand\xviipt{\edef\f@size{\@xviipt}\rm}
\newcommand\xxpt  {\edef\f@size{\@xxpt}\rm}
\newcommand\xxvpt {\edef\f@size{\@xxvpt}\rm}
\makeatother
\usepackage{rmh30,psfig}
\usepackage{amsmath,amsfonts,graphicx,subfigure}
\usepackage[makeindex]{imakeidx}
\usepackage{color,hyperref}
\newfont{\eurbx}{eurb10 at 10pt}
\newfont{\eurbvii}{eurb7 at 7pt}
\newfont{\eurbv}{eurb5 at 5pt}
\newfont{\cmmibx}{cmmib10 at 10pt}
\newfont{\cmmibvii}{cmmib10 at 7pt}
\newfont{\cmmibv}{cmmib10 at 5pt}
\newfont{\eurmx}{eurm10 at 10pt}
\newfont{\eurmvii}{eurm7 at 7pt}
\newfont{\eurmv}{eurm5 at 5pt}
\newfont{\msbmx}{msbm10 at 10pt}
\newfont{\msbmvii}{msbm10 at 7pt}
\newfont{\msbmv}{msbm10 at 5pt}
\newfont{\cmrx}{cmr10 at 10pt}
\newfont{\cmrvii}{cmr7 at 7pt}
\newfont{\cmrv}{cmr5 at 5pt}
\newfont{\cmbxx}{cmbx10 at 10pt}
\newfont{\cmbxvii}{cmbx7 at 7pt}
\newfont{\cmbxiv}{cmbx5 at 5pt}

\newcommand{\chapHelly}{4}

\newcommand{\chapDiscrepancy}{13}

\newcommand{\chapArrg}{28}

\newcommand{\chapRange}{40}

\newcommand{\chapRandomization}{44}

\newcommand{\chapCoreset}{48}

\makeindex
\makeindex[name=term,title=\textsf{INDEX OF DEFINED TERMS},options=-s indexstyle]
\makeindex[name=auth,title=\textsf{INDEX OF CITED AUTHORS},options=-s indexstyle]
\newcommand{\gglos}[1]{\index[auth]{#1}}
\newcommand{\tindex}[1]{\index[term]{#1}}

\def\R{\ensuremath{\mathcal{R}}}
\def\P{\ensuremath{\mathcal{P}}}
\def\Ne{\ensuremath{\mathbb{Ne}}}

\def\S{\ensuremath{\mathcal{S}}}
\def\C{\ensuremath{\mathcal{C}}}
\def\H{\ensuremath{\mathcal{H}}}
\def\Re{\ensuremath{\mathbb{R}}}
\def\k{\ensuremath{\kappa}}
\def\eps{\ensuremath{\epsilon}}
\def\F{\ensuremath{\mathcal{F}}}
\def\T{\ensuremath{\mathcal{T}}}

\DeclareMathOperator{\disc}{\mathrm disc}
\DeclareMathOperator{\vcdim}{\mathrm VC-dim}
\DeclareMathOperator{\opt}{\mathrm OPT}
\DeclareMathOperator{\HD}{\mathrm HD}
\DeclareMathOperator{\dist}{\mathrm dist}
\DeclareMathOperator{\vol}{\mathrm vol}

\newcommand{\remove}[1]{}

\newcommand{\etal}[1]{\emph{et al.\ }}
\makeatletter

\newcommand{\observation}[2]{\stepcounter{thm}
\par\addvspace{12pt}
\noindent{%
\edef\@currentlabel{\thechapter.\theacount.\thethm}
{\hvbxi OBSERVATION \thechapter.\theacount.\thethm}\hspace*{1em}{\trmitxi
#1}\par\addvspace{3pt}\noindent{\trmitx #2}\par\addvspace{3pt}}}

\newcommand{\conj}[2]{\stepcounter{thm}
\par\addvspace{12pt}
\noindent{%
\edef\@currentlabel{\thechapter.\theacount.\thethm}
{\hvbxi CONJECTURE \thechapter.\theacount.\thethm}\hspace*{1em}{\trmitxi
#1}\par\addvspace{3pt}\noindent{\trmitx #2}\par\addvspace{3pt}}}

\newcommand{\lemma}[2]{\stepcounter{thm}
\par\addvspace{12pt}
\noindent{%
\edef\@currentlabel{\thechapter.\theacount.\thethm}
{\hvbxi LEMMA \thechapter.\theacount.\thethm}\hspace*{1em}{\trmitxi
#1}\par\addvspace{3pt}\noindent{\trmitx #2}\par\addvspace{3pt}}}

\newcommand{\theorem}[2]{\stepcounter{thm}
\par\addvspace{12pt}
\noindent{%
\edef\@currentlabel{\thechapter.\theacount.\thethm}
{\hvbxi THEOREM \thechapter.\theacount.\thethm}\hspace*{1em}{\trmitxi
#1}\par\addvspace{3pt}\noindent{\trmitx #2}\par\addvspace{3pt}}}

\newcommand{\corollary}[2]{\stepcounter{thm}
\par\addvspace{12pt}
\noindent{%
\edef\@currentlabel{\thechapter.\theacount.\thethm}
{\hvbxi COROLLARY \thechapter.\theacount.\thethm}\hspace*{1em}{\trmitxi
#1}\par\addvspace{3pt}\noindent{\trmitx #2}\par\addvspace{3pt}}}

\makeatother

\begin{document}
\setcounter{chapter}{47}
\setcounter{section}{0}
\setcounter{page}{1241}
\copyrightdata{978-1-4987-1139-5}{\copyright 2017
by CRC Press, Inc.}

\runhds{N.H.~Mustafa and K.~Varadarajan}
{Chapter \thechapter: $\epsilon$-approximations and $\epsilon$-nets}

\thispagestyle{plain}

\vspace{-1pc}

\noindent\hskip-3pc{\hvbxxiv \thechapter\hskip 18pt EPSILON-APPROXIMATIONS \& EPSILON-NETS}

\vspace{1pc}

\noindent{\hvxiv Nabil H. Mustafa\footnote{The work of Nabil H. Mustafa
 has been supported by the grant ANR SAGA (JCJC-14-CE25-0016-01).} and Kasturi Varadarajan}

\vspace{4pc}

\textheight=50pc

\Bnn{INTRODUCTION}

\noindent The use of random samples to approximate
properties of geometric configurations
has been an influential idea  for both combinatorial and algorithmic purposes.
This chapter considers  two related notions---$\eps$-approximations
and $\eps$-nets---that capture the most important quantitative
properties that one would expect from a random sample with
respect to an underlying geometric configuration.
An example problem: given a set $P$ of points in the plane
and a parameter $\eps>0$, is it possible to choose a
set $N$ of $O(\frac{1}{\eps})$ points of $P$ such that $N$ contains
at least one point from \emph{each} disk containing $\eps |P|$ points of $P$?
More generally, what is the smallest non-empty set $A \subseteq P$
that can be chosen such that for any disk $D$ in the plane,
the proportion of points of $P$ contained in $D$ is within $\eps$
to the proportion of points of $A$ contained in $D$?
In both these cases, a random sample provides an answer
``in expectation''; establishing worst-case guarantees
is the topic of this chapter.

%After considering combinatorial properties
%of commonly studied geometric configurations in Section 1.1,
%the chapter gives a detailed

\A{SET SYSTEMS DERIVED FROM GEOMETRIC CONFIGURATIONS}

\noindent Before we present work on $\eps$-approximations
and $\eps$-nets
for geometric set systems,
we   briefly survey
different types of set systems that can be derived
from geometric configurations and
study the combinatorial properties of these
set systems due to the constraints
induced by geometry.
For example, consider the fact
that for any set $P$ of points
in the plane, there are only $O(|P|^3)$ subsets of $P$ induced by containment
by disks. This is an immediate consequence
of the property that three points of $P$ are
sufficient to ``anchor'' a disk.
This property will be abstracted to a purely combinatorial one,
called the \emph{VC-dimension} of a set system, from which can be derived many
analogous properties for abstract set systems.

\Bnn{GLOSSARY}

\begin{gllist}
\item {\tindex{set system}\trmbitx Set systems:}\quad A pair
%\todo{do we really need to use the notation of $\Sigma$?}
$\Sigma = (X, \R)$, where $X$ is a set of base elements
and $\R$ is a collection of subsets of $X$, is called
a set system. The \emph{dual set system} to  $(X, \R)$
is the system $\Sigma^* = (X^*, \R^*)$, where $X^* = \R$,
and for each $x \in X$, the set $\R_x := \{ R \in \R : x \in R\}$
belongs to $\R^*$.

\item {\tindex{VC-dimension}\trmbitx VC-dimension:}\quad
For any set system $(X, \R)$ and $Y \subseteq X$, the projection
of $\R$ on $Y$ is the set system
$\R|_{Y} := \big\{Y \cap R: R\in \R \big\}$.
The Vapnik-Chervonenkis dimension (or VC-dimension)
of $(X, \R)$, denoted as $\vcdim(\R)$, is the minimum integer $d$ such that $|\R|_{Y}| < 2^{|Y|}$ for any finite subset $Y\subseteq X$ with $|Y| > d$.

\item {\tindex{shatter function!of a set system}\trmbitx Shatter function:}\quad
A set $Y$ is \emph{shattered} by $\R$ if $|\R|_Y| = 2^{|Y|}$.
The \emph{shatter function}, $\pi_{\R}: \Ne \rightarrow \Ne$,
of a set system $(X, \R)$
is obtained by letting $\pi_{\R}(m)$ be the maximum number of subsets
in $\R|_Y$ for any set $Y \subseteq X$ of size $m$.

\item {\tindex{shallow-cell complexity!of a set system}\trmbitx Shallow-cell complexity:}\quad
\ \ A set system $(X, \R)$ has \emph{shallow-cell complexity}
${\varphi_{\R}:\Ne \times \Ne \to \Ne}$,  if for every $Y \subseteq X$,
the number of sets of size at most $l$ in the set system $\R|_Y$ is $O \big( |Y| \cdot \varphi_{\R}(|Y|, l)  \big)$. For convenience, dropping
the second argument of $\varphi_{\R}$, we say that
$(X,\R)$ has \emph{shallow-cell complexity}
$\varphi_{\R}:\Ne \to \Ne$, if there exists a constant $c(\R)>0$ such that for every $Y \subseteq X$ and for every positive integer $l$, the number of sets
of size at most $l$ in $\R|_Y$ is $O \big( |Y| \cdot \varphi_{\R}(|Y|) \cdot l^{c(\R)} \big)$.

\item {\tindex{set system!geometric}\trmbitx Geometric set systems:}\quad
Let $\R$ be a family of (possibly unbounded) geometric objects in $\Re^d$,
and $X  $ be a finite set of points in $\Re^d$.
%\todo{primal need not involve points but geometric objects.} Then the set system
Then the set system
$(X, \R|_X)$ is called a  \emph{primal set system induced by $\R$}.
Given a finite set $\S \subseteq \R$, the
\emph{dual set system induced by $\S$} is  the set system
$(\S, \S^*)$, where $\S^* = \{S_x \ : \ x \in \Re^d\}$
and $S_x := \big\{S \in \S \ : \ x \in S\big\}$.
%\todo{how to denote the dual set system induced by $\R$}

\item {\tindex{union complexity} \trmbitx Union complexity of geometric objects:}\quad
The {\em union complexity}, $\kappa_{\R}: \Ne \to \Ne$, of a family of objects $\R$ is
obtained by letting $\kappa_{\R}(m)$ be the maximum number of faces of all dimensions that the union of any $m$ members of $\R$ can have.

\item {\tindex{set system!delta-separated@$\delta$-separated} \trmbitx $\delta$-Separated set systems:}\quad
The symmetric difference of two sets $R, R'$ is denoted as $\Delta(R,R')$,
where $\Delta(R, R') = (R \setminus R' ) \cup (R' \setminus R)$.
Call a set system $(X, \R)$  $\delta$-separated if
for every pair of sets $R, R' \in \R$, $|\Delta(R, R')| \geq \delta$.

%\item {\tindex{Veronese maps} \trmbitx Veronese maps:}\quad

\end{gllist}

\vspace{-.8pc}

\Bnn{VC-DIMENSION}

\noindent
First defined by Vapnik and Chervonenkis~\cite{VC71}, a crucial
property of VC-dimension is that it is \emph{hereditary}---if
a set system $(X, \R)$ has VC-dimension $d$,
then for any $Y \subseteq X$, the VC-dimension of
the set system $(Y, \R|_Y)$ is at most $d$.

\vspace{-.2pc}
\lemma{\cmrx\cite{VC71,Sa72,Sh72}\label{lemma:saeurshelah}}
{Let $(X, \R)$ be a set system
with $\vcdim(\R) \leq d$ for a fixed constant $d$. Then
for all positive integers $m$,
$$\pi_{\R}(m) \leq \sum_{i=0}^d \binom{m}{i}
= O\left( \left(\frac{em}{d} \right)^d \right).$$
Conversely,
if $\pi_{\R}(m) \leq cm^d$ for some constant $c$, then $\vcdim(\R) \leq 4 d \log (cd)$.}
%\dtodo{precise quantitative relation?}

Throughout this chapter, we usually state the results
in terms of   shatter functions of  set systems; the
first part of Lemma~\ref{lemma:saeurshelah}  implies
that these results carry over for set systems with bounded
VC-dimension as well.
Geometric set systems often have bounded VC-dimension, a key
case being the primal set system induced by half-spaces
in $\Re^d$, for which Radon's lemma~\cite{R21} implies the following.

\vspace{-.2pc}
\lemma{\cmrx\label{lemma:halfspacesVCdim}}
{Let $\H$ be the family of all half-spaces in $\Re^d$.
Then  $\vcdim(\H) = d+1$. Consequently, $\pi_{\H}(m) = O(m^{d+1})$.}

Lemma~\ref{lemma:halfspacesVCdim} is the starting point
for bounding the VC-dimension of a large category of geometric set systems.
For example, it implies
that the VC-dimension of the primal set system induced
by balls in $\Re^{d}$ is $d+1$, since
if a set  of points  is shattered by
the primal set system induced by balls, then it is also shattered
by the primal set system induced by half-spaces\footnote{Assume
that a set $X$ of  points
in $\Re^d$ is shattered by the primal set system induced by balls.
Then for any $Y \subseteq X$, there exists a ball $B$ with
$Y = B \cap X$, and a ball $B'$ with $X \setminus Y = B' \cap X$.
Then any hyperplane that separates $B \setminus B'$  from $B' \setminus B$
also separates
$Y$ from $X \setminus Y$.}.
More generally,  sets defined by
polynomial  inequalities can be lifted
to half-spaces in some higher dimension by Veronese maps and so
also have bounded VC-dimension. Specifically, identify
each $d$-variate polynomial $f(x_1, \ldots, x_d)$
with its induced set $S_f := \big\{ p \in \Re^d: f(p) \geq 0 \big\}$. Then
Veronese maps---i.e., identifying
the $d' = \binom{D+d}{d}$ coefficients
of a $d$-variate polynomial of degree at most $D$
with   distinct coordinates of $\Re^{d'}$---together with Lemma~\ref{lemma:halfspacesVCdim} immediately
imply the  following.

\lemma{\cmrx\cite{MatLectures}}
{Let $\R_{d,D}$ be the primal set system
induced by all $d$-variate polynomials over $\Re^d$ of degree at most $D$.
Then $\vcdim(\R_{d,D}) \leq \binom{D+d}{d}$.}

%The above bound can be improved via direct arguments for
%specific cases.

Set systems derived from other  bounded VC-dimension set systems
using a finite sequence of set
operations  can be shown
to also have bounded VC-dimension. The number of
sets in this derived set system can be computed by
a direct combinatorial argument, which together
with the second part of Lemma~\ref{lemma:saeurshelah}
implies the following.

\lemma{\cmrx\cite{HW87}}
{Let $(X, \R)$ be a set system with ${\vcdim(\R) \leq d}$,
and $k \geq 1$ an integer.
Define the set system
$$F_k(\R) := \big\{ F(R_1, \ldots, R_k): R_1, \ldots, R_k \in \R\big\},$$
where $F(S_1, \ldots, S_k)$ denotes the set derived
from the input sets $S_1, \ldots, S_k$ from a fixed finite
sequence of union, intersection and difference operations.
Then we have ${\vcdim \big(F_k(\R)\big) = O(kd \log k)}$.}

%There is a close relation between VC-dimension of primal and
%dual set systems.

%\vspace{-1pc}

\lemma{\cmrx\cite{A83} \label{lemma:dualvcdim}}
{Given a set system $\Sigma = (X, \R)$ and
its dual system $\Sigma^* = (X^*, \R^*)$,
$\vcdim(\R^*) < 2^{\vcdim(\R)+1}$.}
Note that if $\pi_{\R^*}(m) = O \left(m^d \right)$ for some constant $d$,
then the second part of Lemma~\ref{lemma:saeurshelah}
implies that $\vcdim \left(\R^* \right) = O(d \log d)$, 
and Lemma~\ref{lemma:dualvcdim} then implies that 
$\vcdim \left( \R \right) = 2^{O ( d \log d)} =  d^{O(d)}$.

On the other hand,   the primal set system
induced by  convex objects
in $\Re^2$ has unbounded VC-dimension,
as it shatters any set of points in convex position.

\Bnn{SHALLOW-CELL COMPLEXITY}
\noindent
A key realization following
from the work of Clarkson and Varadarajan~\cite{CV07} and Varadarajan~\cite{V10}
was to consider a finer classification
of set systems than just based on VC-dimension, namely its shallow-cell complexity, first defined explicitly in Chan \etal~\cite{CGKS12}.
Note that if $(X, \R)$ has shallow-cell complexity $\varphi_{\R}(m) = O(m^t)$ for some constant $t$,
then $\pi_{\R}(m) = O\big(m^{1+t+c(\R)}\big)$ for an absolute
constant $c(\R)$, and so $\R$ has bounded
VC-dimension.
On the other hand, while the shatter function bounds the total number of sets
in the projection of $\R$ onto a subset $Y$, it does not give
any information on the \emph{distribution} of the set sizes,
which has turned out to be a key parameter (as we will see
later in, e.g., Theorem~\ref{theorem:shallowcellcomplexityepsnets}).
Tight bounds on shatter functions and shallow-cell complexity
are known for many basic geometric set systems.
%For instance, the bounds on $k$-sets for half-spaces and
%union complexity of objects in $\Re^2$ imply the following
%statements.

\lemma{\cmrx\cite{CS89}\tindex{k-set@$k$-set!for half-spaces}\label{lemma:shallowcellforhalfspaces}}
{Let $\H$ be the family of all half-spaces
in $\Re^d$. Then $\varphi_{\H}(m) = O(m^{\lfloor d/2 \rfloor-1})$.
Furthermore, this bound is tight, in the sense that for any integer $m \geq 1$,
there exist $m$ points for which the above bound can be attained.}

The following lemma, a
%well-known
consequence of a probabilistic technique by Clarkson and Shor~\cite{CS89}, bounds the shallow-cell complexity of the dual set system induced by a set of objects in $\Re^2$.

\lemma{\cmrx\cite{S91}}
{Let $\R$ be a finite set of objects in $\Re^2$,
each bounded by a closed Jordan curve, and with union complexity
$\k_{\R}(\cdot)$.
Further, each intersection point in the arrangement
of $\R$ is defined by a constant number of objects of $\R$.
Then the shallow-cell complexity
of the dual set system induced by $\R$ is bounded by
 $\varphi_{\R^*}(m) = O\big(\frac{\k_{\R}(m)}{m}\big)$.}

Table~\ref{table:shallowcellcomplexity} states the
shatter function  as well as the shallow-cell
complexity of some commonly used set systems.
Some of these bounds are derived from the above two lemmas using
known bounds on union complexity of geometric objects
(e.g., pseudo-disks~\cite{BPR13}, fat triangles~\cite{AdBES14}).
%while others are via   explicit constructions~\cite{NT09}.
%\todo{other examples}.

\begin{table}[t]
\viiipt
\baselineskip=11pt
\renewcommand{\arraystretch}{.9}
\Table{Combinatorial properties of some primal (P)
and dual (D) geometric set systems.}
{\begin{tabular}{| l | l | c | c | c |}
    \hline
\rule[-4pt]{0pt}{13pt}{\hvbviii OBJECTS  }
    & {\hvbviii SETS}
    & {\hvbviii  $\varphi(m)$}
    & {\hvbviii $\vcdim$}
    & {\hvbviii $\pi(m)$}
    \\ \hline
\rule[0pt]{0pt}{9pt}Intervals
\label{table:shallowcellcomplexity}
    & P/D
    & $O(1)$
    & $2$
    & $\Theta(m^2)$
    \\
Lines in $\Re^2$
    & P/D
    & $O(m)$
    & $2$
    & $\Theta(m^2)$
    \\
Pseudo-disks in $\Re^2$
    & P
    & $O(1)$
    & $3$
    & $O(m^{3})$
    \\
Pseudo-disks in $\Re^2$
    & D
    & $O(1)$
    & $O(1)$
    & $O(m^{2})$
    \\
Half-spaces in $\Re^d$
    & P/D
    & $O\big(m^{\lfloor d/2 \rfloor -1}\big)$
    & $d+1$
    & $\Theta(m^d)$
    \\
Balls in $\Re^d$
    & P
    & $O\big(m^{\lceil d/2 \rceil -1}\big)$
    & $d+1$
    & $\Theta(m^{d+1})$
    \\
Balls in $\Re^d$
    & D
    & $O\big(m^{\lceil d/2 \rceil -1}\big)$
    & $d+1$
    & $\Theta(m^{d})$
    \\
Triangles in $\Re^2$
    & D
    & $O(m)$
    & $7$
    & $O(m^7)$
    \\
Fat triangles in $\Re^2$
    & D
    & $O(\log^* m)$
    & $7$
    & $O(m^7)$
    \\
Axis-par. rect. in $\Re^2$
    & P
    & $O(m)$
    & $4$
    & $\Theta(m^4)$
    \\
Axis-par. rect. in $\Re^2$
    & D
    & $O(m)$
    & $4$
    & $\Theta(m^2)$
    \\
Convex sets in $\Re^d$
    & P
    & $O\big(2^m/m \big)$
    & $\infty$
    & $\Theta(2^m)$
    \\
Translates of a convex set in $\Re^d, d \geq 3$
    & P
    & $O\big(2^m/m \big)$
    & $\infty$
    & $\Theta(2^m)$
    \\
    \hline
\end{tabular}}
\renewcommand{\arraystretch}{1}
\xpt
\baselineskip=12pt
\end{table}

\smallskip\noindent{\trmbitx A packing lemma.}
A key combinatorial statement at the heart
of many of the results in this chapter is inspired by packing properties
of geometric objects. It was first proved
for the primal set system induced by half-spaces in $\Re^d$
by geometric techniques~\cite{CW89};
the following more general form was first shown
by Haussler~\cite{H92}\footnote{The theorem as stated in~\cite{H92}
originally required that $\vcdim(\P) \leq d$.
It was later verified that   the proof also works
with the assumption of   polynomially bounded
shatter functions; see~\cite{M95} for details.}
(see~\cite[Chapter 5.3]{MatBook} for a nice exposition of this result).
%, and finally
%generalized to the following form in Mustafa~\cite{M16}.
%
%\begin{lemma}[\cite{H92, C92, M16}]
%Let $(X, \P)$ be a $\delta$-separated set system with $\vcdim(\R) \leq d$. Then
%$$ |\P| \leq 2 \cdot \Ex \big[ | \P|_{S} | \big], \text{where $S \subseteq X$ is a uniform random sample of size $\frac{4dn}{\delta}-1$}.$$
%\end{lemma}
%%\todo{give a short insight into the proof.}
%By bounding the sizes of $\P|_S$ using the shatter function:

\lemma{\cmrx\cite{H92}\tindex{packing lemma}\label{lem:packing}}
{Let $(X, \P)$, $|X|=n$, be a $\delta$-separated set system with $\delta \geq 1$
and   $\pi_{\P}(m) = O(m^d)$ for some constant $d > 1$.
Then $|\P| \leq e \left( d + 1 \right) \left( \frac{2en}{\delta} \right)^d = O\left( \left(\frac{n}{\delta} \right)^d \right)$. Furthermore, this bound is asymptotically tight.}

A strengthening of this statement, for specific values of $\delta$,
was studied for some
geometric set systems  in~\cite{PR08, MR14}, and
  for any $\delta\geq 1$ for the so-called Clarkson-Shor set systems in~\cite{E14, DEG15}.
This was then generalized in terms
of the shallow-cell complexity of a set system to give
the following statement.

% was presented in~\cite{DEG15}.
%%for the case
%%where shallow-cell complexity is polynomially bounded,
%This was then further generalized in~\cite{M16}.

\lemma{\cmrx\cite{M16}\tindex{packing lemma!shallow}\label{lem:shallowpacking}}
{Let $(X, \P)$, $|X|=n$, be a $\delta$-separated set system with $\pi_{\P}(m) = O(m^d)$
for some constant $d > 1$,
and with shallow-cell complexity $\varphi_{\P}(\cdot, \cdot)$.
If $|P| \leq k$ for all $P \in \P$, then $|\P| \leq   O\big( \frac{n}{\delta} \cdot \varphi_{\P} (  \frac{4dn}{\delta}, \frac{24dk}{\delta}) \big)$.}

\noindent A matching lower-bound for Clarkson-Shor set systems was given
in~\cite{DGJM16}.

%\vspace{-0.8pc}

\A{EPSILON-APPROXIMATIONS}

\noindent
Given a set system $(X, \R)$ and a set $A \subseteq X$, a set $R \in \R$ is
\emph{well-represented} in $A$ if $\frac{|R|}{|X|} \approx \frac{|R \cap A|}{|A|}$.
Intuitively, a set $A \subseteq X$ is an $\eps$-approximation for $\R$
if {\em every} $R \in \R$ is well-represented in $A$;
the parameter $\eps$ captures quantitatively the additive error between
these two quantities. In this case the value $\frac{|R \cap A| }{|A|} \cdot |X|$ is a good estimate for $|R|$. As an example, suppose that $X$ is a finite set of points in the plane, and let $A$ be an $\eps$-approximation for the primal set system  on $X$ induced by   half-spaces. Then given a query half-space $h$, one can return
$\frac{|h \cap A| }{|A|} \cdot |X| $ as an estimate for $|h \cap X|$. If $|A| \ll |X|$, computing this estimate is more efficient than   computing $|h \cap X|$.

\Bnn{GLOSSARY}

\begin{gllist}
\item {\tindex{epsilon-approximation@$\varepsilon$-approximation}\trmbitx $\eps$-Approximation:}\quad
Given a finite set system $(X, \R)$, and a parameter $0 \leq \eps \leq 1$,
a set $A \subseteq X$ is called an $\eps$-approximation if,
for each $R \in \R$,
$$ \left|  \frac{|R|}{|X|} - \frac{|R \cap A|}{|A|}  \right| \leq \eps.  $$

\item {\tindex{epsilon-approximation@$\varepsilon$-approximation!sensitive} \trmbitx Sensitive $\epsilon$-approximation:}\quad
Given a set system $(X, \R)$ and a parameter $0 < \epsilon \leq 1$,
a set $A \subseteq X$ is a sensitive $\epsilon$-approximation if for each
$R \in \R$,
\[ \left| \frac{|R|}{|X|} - \frac{|R \cap A|}{|A|} \right| \leq \frac{\epsilon}{2}
\Bigg( \sqrt{\frac{|R|}{|X|}} + \eps \Bigg). \]

\item {\tindex{epsilon-approximation@$\varepsilon$-approximation!relative} \trmbitx Relative $(\epsilon, \delta)$-approximation:}\quad
Given a set system $(X, \R)$ and parameters $0 < \delta, \epsilon \leq 1$,
a set $A \subseteq X$ is a relative $(\epsilon, \delta)$-approximation if
for each $R \in \R$,
%\begin{eqnarray*}
%\Big| \frac{|R|}{|X|} - \frac{|R \cap A|}{|A|} \Big|   \leq
%\begin{cases}
%  \delta  \frac{|R|}{|X|} \text{ \ \ \ \ if } |R| \geq \eps |X|\\
%  \delta \cdot \eps \text{ \ \ \ \ \ otherwise. }
%\end{cases}
%\end{eqnarray*}
\begin{eqnarray*}
\left| \frac{|R|}{|X|} - \frac{|R \cap A|}{|A|} \right|   \leq
\max \Big\{ \delta \cdot \frac{|R|}{|X|}, \ \delta \cdot \eps \Big\}
\end{eqnarray*}

\item {\tindex{discrepancy!combinatorial}\trmbitx Discrepancy:}\quad
Given a set system $(X, \R)$, and a two-coloring $\chi: X \rightarrow \{-1,1\}$,
  define the discrepancy of $R \in \R$ with respect to $\chi$ as $\disc_{\chi}(R) = \big| \sum_{p \in R} \chi(p)  \big|$, and the discrepancy of $\R$ with respect to $\chi$  as
$\disc_{\chi}(\R) = \max_{R \in \R} \disc_{\chi}(R)$. The discrepancy
of $(X, \R)$ is $\disc(\R) = \min_{\chi: X \rightarrow \{-1,1\}}  \disc_{\chi}(\R)$.
\end{gllist}

\Bnnnr{EPSILON-APPROXIMATIONS AND DISCREPANCY}
%\Bnn{EPSILON-APPROXIMATIONS AND DISCREPANCY}

\noindent
When no other constraints are known for a given set system $(X, \R)$,
  the following is the currently best bound on the sizes of
$\eps$-approximations for $\R$.

\theorem{\cmrx\cite{Cha00}\label{thm:epsapprox-general}}
{Given a finite set system $(X, \R)$ and a parameter
${0 < \epsilon \leq 1}$, an $\epsilon$-approximation for $(X, \R)$ of size $O\big(\frac{1}{\eps^2} \log |\R| \big)$ can be found in deterministic $O\big( |X| \cdot |\R| \big)$ time.}

If $\vcdim(\R) = d$, the shatter function $\pi_{\R}(m)$ for $(X, \R)$ is bounded by $O(m^d)$ (Lemma~\ref{lemma:saeurshelah}). In this case, $|\R| = O\big(|X|^d\big)$, and Theorem~\ref{thm:epsapprox-general} guarantees an $\epsilon$-approximation of size at most $O\big(\frac{d}{\eps^2} \log |X| \big)$. An influential idea originating in the work
of Vapnik and Chervonenkis~\cite{VC71} is that for any set system  $(X, \R)$
with $\vcdim(\R) \leq d$, one can construct an $\eps$-approximation
of $\R$ by uniformly sampling  a subset  $A \subseteq X$ of
size $O\big( \frac{d \log \frac{1}{\eps}}{\eps^2} \big)$.
%A beautiful result of Vapnik and Chervonenkis \cite{VC71} provides an affirmative answer.
Remarkably, this gives a bound on sizes of $\eps$-approximations which
are independent of $|X|$ or $|\R|$.
To get an idea behind the proof, it should  be first noted
that the factor of $\log |\R|$ in Theorem~\ref{thm:epsapprox-general}
comes from applying union bound to a number of failure events,
one for each set in $\R$.
The key idea
in the proof of~\cite{VC71}, called \emph{symmetrization}, is to
``cluster'' failure events based
on comparing the random sample $A$ with
a second sample (sometimes called a
\emph{ghost sample} in   learning theory literature; see~\cite{DGL96}).
Together
with later work which removed the logarithmic
factor,
one arrives at the following.

\vspace{-0.1in}

\theorem{\cmrx\cite{VC71,T94,LLS01}\tindex{epsilon-net theorem@$\varepsilon$-net theorem}\label{thm:epsapprox-VC}}
{Let $(X, \R)$ be a finite set system with $\pi_{\R}(m) = O(m^d)$
for a constant $d \geq 1$, and $0 < \epsilon, \gamma < 1$ be given parameters. Let $A \subseteq X$ be a subset of size
\[ c \cdot \left( \frac{ d }{\eps^2} + \frac{\log \frac{1}{\gamma}}{\eps^2} \right) \]
chosen uniformly at random, where $c$ is a sufficiently large constant. Then $A$ is an $\epsilon$-approximation for $(X, \R)$ with probability at least $1 - \gamma$.}
%\todo{http://blog.geomblog.org/2011/02/sample-complexity-for-eps.html unable to understand Talagrand's paper}

The above theorem immediately implies a randomized algorithm for computing
approximations. There   exist  near-linear time
deterministic algorithms for constructing
$\eps$-approximations of size slightly worse than the above bound;
see~\cite{STZ06} for algorithms for computing $\eps$-approximations
in data streams.

\theorem{\cmrx\cite{CM96}\label{thm:epsapprox-VC-runningtime}}
{Let $(X, \R)$ be a  set system with ${\vcdim(\R) = d}$, and $0 < \epsilon \leq \frac{1}{2}$ be a given parameter. Assume that given any finite $Y \subseteq X$, all the sets
in $\R|_Y$ can be computed explicitly in time $O\big(|Y|^{d+1}\big)$.
Then an $\eps$-approximation for $(X, \R)$
of size $O\big( \frac{d}{\eps^2} \log \frac{d}{\eps} \big)$ can be computed deterministically
in  $O\big( d^{3d} \big) \big( \frac{1}{\eps^2} \log \frac{d}{\eps} \big)^d|X|$ time.}

Somewhat surprisingly, it is possible to show the
existence of $\epsilon$-approximations of size smaller
than that guaranteed by Theorem~\ref{thm:epsapprox-VC}.
Such results are usually established using
a fundamental relation between the notions of approximations
and discrepancy: assume $|X|$ is even and
let $\chi: X \rightarrow \{-1,+1\}$ be any two-coloring of $X$.
For any $R \subseteq X$, let $R^+$ and $R^-$ denote
the subsets of $R$ of the two colors, and
w.l.o.g., assume that
$|X^+| = \frac{|X|}{2} + t$ and $|X^-| = \frac{|X|}{2} - t$
for some integer $t \geq 0$.
Assuming that $X \in \R$, we have
$\big| |X^+| - |X^-| \big| \leq   \disc_{\chi}(\R)$,
and so $t \leq \frac{\disc_{\chi}(\R)}{2}$.
Take $A$ to be any subset of $X^+$ of size $\frac{|X|}{2}$.
Then for any $R \in \R$,
$$\big| |R^+| - |R^-| \big| =
\big| |R^+| -  ( |R| - |R^+|  ) \big|
  \leq \disc_{\chi}(\R) \implies \Big| |R^+| - \frac{|R|}{2} \Big| \leq \frac{\disc_{\chi}(\R)}{2}.$$
 As $|R \cap A| \geq |R^+| - t$,
 this implies that $\Big| |R \cap A| - \frac{|R|}{2} \Big| \leq \disc_{\chi}(\R)$. Thus
$$ \Big|  \frac{|R|}{|X|} - \frac{|R \cap A|}{|A|}  \Big| \leq
 \Big|  \frac{|R|}{|X|} - \frac{\frac{|R|}{2} \pm \disc_{\chi}(\R)}{\frac{|X|}{2}} \Big| \leq \frac{2 \cdot \disc_{\chi}(\R)}{|X|}, $$
and we arrive at the following.

\lemma{\cmrx\cite{MWW93}\label{lem:disc-epsapprox}}
{Let $(X,\R)$ be a set system with $X \in \R$, and let $\chi: X \rightarrow \{+1, -1 \}$ be any two-coloring of $X$. Then there exists
a set $A \subset X$, with $|A| = \lceil \frac{|X|}{2} \rceil$, such that $A$ is an $\epsilon$-approximation for $(X,\R)$, with $\epsilon = \frac{2 \cdot \disc_{\chi}(\R)}{|X|}$.}

The following simple observation on $\epsilon$-approximations is quite useful.
%; for a proof, see for instance \cite{MatBook}.
\observation{\cmrx\cite{MWW93}\label{obs:epsapprox-compose}}
{If $A$ is an $\epsilon$-approximation for $(X, \R)$, then any $\epsilon'$-approximation for $(A, \R|_{A})$ is an $(\epsilon + \epsilon')$-approximation for
$(X,\R)$.}

Given a finite set system $(X, \R)$ with $X \in \R$,
put  $X_0 = X$, and compute a sequence $X_1, X_2, \ldots, X_t$, where $X_i
\subseteq X_{i-1}$ satisfies $|X_i| = \Big\lceil \frac{|X_{i-1}|}{2} \Big\rceil$, and is computed from a two-coloring of $(X_{i-1}, \R|_{X_{i-1}})$ derived
from Lemma~\ref{lem:disc-epsapprox}. Assume that $X_i$ is an $\eps_i$-approximation for $(X_{i-1}, \R|_{X_{i-1}})$. Then Observation~\ref{obs:epsapprox-compose} implies that $X_t$ is a $\eps$-approximation for $(X,\R)$ with $\eps = \sum_{i=1}^t \epsilon_i$.  The next
statement follows
%from computing an $\epsilon$-approximation of $(X,\R)$
by setting the parameter $t$ to be as large as possible while ensuring that $\sum_{i=1}^t \epsilon_i \leq \epsilon$.

\lemma{\cmrx\cite{MWW93}\label{lem:epsapprox-via-disc}}
{Let $(X,\R)$ be a finite set system with $X \in \R$, and let $f(\cdot)$ be a function such that $\disc \big( \R|_Y \big) \leq f\big(|Y|\big)$ for all $Y \subseteq X$. Then, for every integer $t \geq 0$, there exists an $\epsilon$-approximation $A$ for $(X, \R)$ with $|A| =   \lceil \frac{n}{2^t} \rceil$ and
\[ \epsilon \leq \frac{2}{n} \Bigg( f(n) + 2 f \Big(\Big\lceil \frac{n}{2} \Big\rceil \Big) + \cdots + 2^{t} f\Big(\Big\lceil \frac{n}{2^{t}} \Big\rceil\Big) \Bigg). \]
In particular, if there exists a constant $c > 1$ such that we have $f(2m) \leq \frac{2}{c} f(m)$ for all $\displaystyle m \geq  \lceil \frac{n}{2^t} \rceil$, then $\displaystyle \epsilon = O\Big( \frac{f\big(\lceil \frac{n}{2^t} \rceil \big) 2^t}{n} \Big)$.}

Many of the currently best bounds on $\eps$-approximations follow
from applications of Lemma~\ref{lem:epsapprox-via-disc};
e.g., the existence of $\eps$-approximations of size
$O\big( \frac{1}{\eps^2} \log \frac{1}{\eps} \big)$
for set systems $(X, \R)$ with $\pi_{\R}(m) = O(m^d)$ (for some constant $d>1$)
follows immediately from the fact that for such $\R$, we have
$\disc(\R|_Y) = O\big( \sqrt{ |Y| \log |Y| } \big)$.
The next two theorems, from a seminal paper of
Matou{\v{s}}ek, Welzl, and Wernisch~\cite{MWW93}, were established
by deriving improved discrepancy bounds (which
turn out to be based on Lemma~\ref{lem:packing}), and
then applying Lemma~\ref{lem:epsapprox-via-disc}.

\theorem{\cmrx\cite{MWW93, M95}\tindex{epsilon-approximation@$\varepsilon$-approximation!for bounded VC-dimension}\label{thm:epsapprox-primal}}
{Let $(X,\R)$ be a finite set system  with the shatter
function $\pi_{\R}(m) = O\big(m^d\big)$, where
$d > 1$ is a fixed constant.
For any $0 < \eps \leq 1$,
there exists
an $\epsilon$-approximation for $\R$ of size $\displaystyle O \left( \frac{1}{\epsilon^{2 - \frac{2}{d+1}}} \right)$.}

The above theorem relies on the existence of
low discrepancy colorings, whose
initial proof was non-algorithmic (using the ``entropy method'').
However, recent work by Bansal~\cite{B12} and Lovett and Meka~\cite{LM15} implies
polynomial time algorithms for constructing such low discrepancy colorings
and consequently $\eps$-approximations whose sizes are given by Theorem~\ref{thm:epsapprox-primal}; see~\cite{E14, DEG15}.

Improved bounds on approximations are also known
in terms of the shatter function of the set system dual to $(X,\R)$.

\theorem{\cmrx\cite{MWW93}\label{thm:epsapprox-dual}}
{Let $(X,\R)$ be a finite set system and $0 < \eps \leq 1$ be a given parameter. Suppose that for the set system $(X^*, \R^*)$ dual to $(X, \R)$, we have $\pi_{\R^*}(m) = O \big(m^d\big)$, where $d > 1$ is a  constant independent of $m$. Then there
%\todo{Matousek uses $\pi^*_{\R}$ instead of $\pi_{\R^*}$}
exists an $\epsilon$-approximation for $\R$ of size $\displaystyle O \left( \frac{1}{\epsilon^{2 - \frac{2}{d+1}}}  \big(\log \frac{1}{\eps}\big)^{1 - \frac{1}{d+1}} \right)$.}

Theorems~\ref{thm:epsapprox-primal} and \ref{thm:epsapprox-dual} yield the best known bounds for several geometric  set systems.
For example, the shatter function (see Table~\ref{table:shallowcellcomplexity}) of the primal set system induced
by half-spaces in $\Re^2$ is $O(m^2)$, and thus
 one obtains $\epsilon$-approximations for it of size $O\big( \frac{1}{\epsilon^{4/3}} \big)$ from Theorem~\ref{thm:epsapprox-primal}.
For the primal set system induced by disks in $\Re^2$,
the shatter function is bounded by $\Theta (m^3)$; Theorem~\ref{thm:epsapprox-primal} then
implies the existence of $\epsilon$-approximations of size $O\big( \frac{1}{\epsilon^{3/2}} \big)$. In this case, it turns out that Theorem~\ref{thm:epsapprox-dual} gives
a better bound: the shatter function of the dual set system is bounded by $O(m^2)$,
and thus there exist $\epsilon$-approximations of size $O\big( \frac{1}{\epsilon^{4/3}} (\log \frac{1}{\eps})^{\frac{2}{3}} \big)$.

Table~\ref{table:epsapprox} states the best known bounds for some common geometric set systems. Observe that for the primal  set system induced by axis-parallel rectangles in $\Re^d$, there exist $\epsilon$-approximations of size
near-linear in $\frac{1}{\eps}$.
%$O \left( \frac{1}{\eps} \cdot (\log^{2d}  \frac{1}{\eps}) \cdot \log^{c_d} (\log \frac{1}{\eps}) \right)$, where $c_d > 0$ is a constant that depends on $d$.

%\vspace{-0.6pc}

\begin{table}[htb]
\viiipt
\baselineskip=11pt
\renewcommand{\arraystretch}{1.3}
\Table{Sizes of $\eps$-approximations for geometric set systems (multiplicative constants omitted for clarity).}
{\begin{tabular}{| l | l | l |}
    \hline
\rule[-4pt]{0pt}{13pt}{\hvbviii Objects}
    & {\hvbviii SETS}
    & {\hvbviii UPPER-BOUND}
    \\ \hline
\rule[0pt]{0pt}{9pt}Intervals
\label{table:epsapprox}
    & Primal
    & $\frac{1}{\eps}$
    \\
Half-spaces in $\Re^d$
    & Primal/Dual
    & $\frac{1}{\epsilon^{2 - \frac{2}{d+1}}}$  \hfill \cite{MWW93,M95}
    \\
Balls in $\Re^d$
    & Primal
    & $\frac{1}{\epsilon^{2 - \frac{2}{d+1}}}  (\log \frac{1}{\eps})^{1 - \frac{1}{d+1}}$ \hfill \cite{MWW93}
    \\
Balls in $\Re^d$
    & Dual
    & $\frac{1}{\epsilon^{2 - \frac{2}{d+1}}}$  \hfill \cite{MWW93,M95}
    \\
Axis-par. rect. in $\Re^d$
    & Primal
    & $\frac{1}{\eps} \cdot (\log^{2d}  \frac{1}{\eps}) \cdot \log^{c_d} (\log \frac{1}{\eps})$ \hfill \cite{Ph08}
    \\
    \hline
\end{tabular}}
\renewcommand{\arraystretch}{1}
\xpt
\baselineskip=12pt
\end{table}

\vspace{-1.4pc}
\Bnn{RELATIVES OF EPSILON-APPROXIMATIONS}

\noindent
It is easy to see that a sensitive $\epsilon$-approximation is  an $\eps$-approximation and an $\epsilon'$-net, for $\eps' > \eps^2$ (see the subsequent section for the definition of $\eps$-nets) simultaneously. This notion was first studied by Br{\"o}nnimann \etal~\cite{BCM}.  The following result improves slightly on their bounds.

\pagebreak

\theorem{\cmrx\cite{BCM,HarBook}\label{thm:sensitive-approx}}
{Let $(X, \R)$ be a finite system
with $\vcdim(\R) \leq d$, where $d$ is a fixed
constant. For a given parameter $0 < \epsilon \leq 1$, let $A \subseteq X$ be a subset of size
\[ \frac{c\cdot d }{\eps^2} \log \frac{d}{\eps} \]
chosen uniformly at random, where $c > 0$ is an absolute constant. Then $A$ is a sensitive $\epsilon$-approximation for $(X, \R)$ with probability at least $\frac{1}{2}$.
Furthermore, assuming that given any $Y \subseteq X$, all the sets
in $\R|_Y$ can be computed explicitly in time $O\big(|Y|^{d+1}\big)$,
 a sensitive $\eps$-approximation
of size $O\big( \frac{d}{\eps^2} \log  \frac{d}{\eps}  \big)$
can be computed deterministically in time $O(d^{3d}) \cdot \frac{1}{\eps^{2d}} (\log \frac{d}{\eps})^d \cdot |X|$.}

On the other hand, a relative $(\eps, \delta)$-approximation
%, first introduced in~\cite{CKMS06},
is both a $\delta$-approximation and an $\eps'$-net,
for any $\eps' > \eps$.
It is easy to see that a $(\epsilon \cdot \delta)$-approximation is a relative $(\eps, \delta)$-approximation.  Thus, using Theorem~\ref{thm:epsapprox-VC}, one obtains a relative $(\eps, \delta)$-approximation of size $O \big( \frac{d}{\epsilon^2 \cdot \delta^2} \big)$. This bound can be   improved to the following.

\theorem{\cmrx\cite{LLS01, HS11}\label{thm:relative-approx}}
{Let $(X, \R)$ be a finite set system with   shatter function $\pi_{\R}(m) = O(m^d)$
for some constant $d$, and $0 < \delta, \epsilon, \gamma \leq 1$ be given parameters. Let $A \subseteq X$ be a subset of size
\[ c \cdot \left( \frac{d \log \frac{1}{\eps} }{\eps \delta^2} + \frac{\log \frac{1}{\gamma}}{\eps \delta^2}  \right) \]
chosen uniformly at random, where $c > 0$ is an absolute constant. Then $A$ is a relative $(\eps,\delta)$-approximation for $(X, \R)$ with probability at least $1-\gamma$.}

A further improvement is possible
on the size of relative $(\eps, \delta)$-approximations
for the primal set system induced by half-spaces in $\Re^2$~\cite{HS11}
and $\Re^3$~\cite{E14}, as well as other bounds
with a better dependency on $\frac{1}{\delta}$ (at the cost of a
worse dependence on $\frac{1}{\eps}$) for systems with small shallow-cell
complexity~\cite{E14, DEG15}.

%Har-Peled and Sharir derive this result from a related result on sampling by  Li, Long, and Srinivasan \cite{}.
%In a very similar manner to Theorem~\ref{thm:epsapprox-primal}, one
%can improve the above bound via improved discrepancy bounds for the corresponding
%set systems.
%\begin{theorem}[\cite{E14}]
%Let $(X, \R)$ be a set system with
%$\varphi_{\R}(m, k) = O\big( m^{d_1} k^{d-d_1}\big)$, where $1 < d_1 \leq d$ are constants. Let $0 < \delta, \eps \leq 1$ be given parameters. Then there exists
%a relative $(\eps, \delta)$-approximation of size
%\[ O\Bigg(   \frac{ \log \frac{1}{\eps \delta} }{ \eps^{\frac{d+d_1}{d+1}} \delta^{\frac{2d}{d+1}} } \Bigg). \]
%\end{theorem}
%

% interesting special cases have been explored by Har-Peled and Sharir~\cite{HS11},

\A{APPLICATIONS OF EPSILON-APPROXIMATIONS}

\noindent
One of the main uses of $\eps$-approximations is in
constructing a small-sized representation or ``sketch'' $A$ of a potentially large set of
elements $X$ with respect to an underlying set system $\R$.
Then data queries from $\R$ on $X$ can instead be performed
on $A$ to get provably approximate answers.
Suppose that we aim to preprocess a finite set $X$ of points in the plane, so that given a query half-space $h$, we can efficiently return an approximation to $| h \cap X|$. For this data structure, one could use an $\epsilon$-approximation $A \subseteq X$
for the set system $(X, \R)$ induced by the set of all half-spaces in $\Re^2$.
Then given a query half-space $h$,   simply return
$\frac{|h \cap A|}{|A|} \cdot |X|$; this answer differs from
$|h \cap X|$ by at most $\epsilon \cdot |X|$. If instead $A$ is a relative
$(\delta, \epsilon)$-approximation, then our answer differs from the true answer by at most $\delta \cdot |h \cap X|$, provided $|h \cap X| \geq \eps |X|$.
Two key properties of approximations useful in applications
are $(a)$ $\frac{|R \cap A|}{|A|}$ approximates $\frac{|R \cap X|}{|X|}$
\emph{simultaneously} for each $R \in \R$, and
$(b)$ $\eps$-approximations exist of size independent of $|X|$ or $|\R|$.
This enables the use of $\eps$-approximations
for computing certain estimators on
geometric data; e.g., a combinatorial median
$q \in \Re^d$ for a point set $X$  can be approximated
by the one for an $\eps$-approximation, which can then
be computed in near-linear time.

\Bnn{GLOSSARY}
\begin{gllist}

\item {\tindex{set systems!product} \trmbitx Product set systems:}\quad
Given   finite set systems $\Sigma_1 = (X_1, \R_1)$ and ${\Sigma_2 = (X_2, \R_2)}$, the
product system $\Sigma_1 \otimes \Sigma_2$ is defined as the system $(X_1 \times X_2, \T)$, where $\T$ consists of all subsets $T \subseteq X_1 \times X_2$ for which the following hold: $(a)$ for any $x_2 \in X_2$, $\{x \in X_1 \ : \ (x,x_2) \in T\} \in \R_1$, and $(b)$ for any $x_1 \in X_1$,   $\{x \in X_2 \ : \ (x_1, x) \in T\} \in \R_2$.

\item{\tindex{centerpoint} \trmbitx Centerpoints:}\quad
Given a set $X$ of $n$ points in  $  \Re^d$, a point $q \in \Re^d$ is said to be a centerpoint for $X$ if any half-space containing $q$   contains at least $\frac{n}{d+1}$ points of $X$; for $\eps > 0$, $q$ is said to be an $\epsilon$-centerpoint if any half-space containing $q$   contains at least $(1 - \eps) \frac{n}{d+1}$ points of $X$. By Helly's theorem, a centerpoint exists for all point sets.

\item {\tindex{shape fitting} \trmbitx Shape fitting:}\quad
A shape fitting problem consists of the triple
$(\Re^d, \F, \dist)$, where $\F$ is a family of non-empty closed
subsets (\emph{shapes}) in $\Re^d$ and $\dist:\ \Re^d \times \Re^d \rightarrow \Re^{+}$ is a continuous, symmetric, positive-definite ({\em distance}) function.
%%satisfying satisfies (a) $\dist(p,q) = 0$ if and only if $p = q$, and (b) $\dist(p,q) = \dist(q,p)$. Each shape
%%$F \in \F$ is a non-empty, closed, subset of $\Re^d$.
The {\em distance} of a
point $p \in \Re^d$ from the shape $F \in \F$ is defined as $\dist(p,F) = \min_{q \in F} \dist(p,q)$.
A finite subset $P \subset \Re^d$ defines an instance
of the shape fitting problem, where the goal is to find a shape $F^* = \arg \min_{F \in \F} \sum_{p \in P} \dist(p,F)$.
%that minimizes $\sum_{p \in P} d(p,F)$ over all $F \in \F$. (We assume $\F$ is such that the minimum is attained.)

\item {\tindex{coreset} \trmbitx $\epsilon$-Coreset:}\quad
Given an instance $P \subset \Re^d$ of a shape fitting problem $(\Re^d, \F, \dist)$, and an $\epsilon \in (0,1)$, an $\epsilon$-coreset of size $s$
is a pair $(S, w)$, where $S \subseteq P$, $|S|=s$, and
$w \ : \ S \rightarrow \Re$ is a weight function such that for any $F \in \F$:
\[ \Big| \sum_{p \in P} \dist(p,F) - \sum_{q \in S} w(q) \cdot \dist(q,F) \Big| \leq
   \epsilon \ \sum_{p \in P} \dist(p,F).\]
%The size of the coreset $(S,w)$ is defined to be $|S|$.

\end{gllist}

\vspace{-0.6pc}

\Bnn{APPROXIMATING GEOMETRIC INFORMATION}
\noindent
One of the main uses of $\eps$-approximations is in
the design of
efficient approximation algorithms for combinatorial queries
on geometric data.
%Suppose that there is a certain type of information that we wish to extract from some geometric data. By computing on an $\epsilon$-approximation of the data instead of the original data, we can sometimes obtain an approximation to the information we seek. If the size of the $\eps$-approximation is smaller than that of the original data, this provides a faster algorithm.
An illustrative example is that of computing a centerpoint of a finite point set $X \subset \Re^d$; the proof of the following lemma is immediate.
%Recall that for any non-empty, finite, point set in $\Re^d$, there is at least one centerpoint.

\lemma{\cmrx\cite{MatCenter}\label{lem:centerpoint}}
{Let $X \subset \Re^d$ be a finite point set, $0 \leq \epsilon < 1$ be a given parameter, and $A$ be an $\eps$-approximation for the primal set system  induced by half-spaces
in $\Re^d$ on $X$. Then any centerpoint for $A$ is an $\epsilon$-centerpoint for $X$.}

We now describe a more subtle application in the same spirit, in fact
one of the motivations for considering products of set systems, first
considered in Br{\"o}nnimann, Chazelle, and Matou{\v{s}}ek~\cite{BCM}.
For $i=1,2$, let $X_i$ be a finite set of lines in $\Re^2$
such that $X_1 \cup X_2$ is in general position,
and let $\R_i$
be the family of subsets of $X_i$
that contains every subset $X'  \subseteq X_i$ such that $X'$ is precisely the subset of lines intersected by some line segment.
 The VC-dimension of the set system $\Sigma_i = (X_i, \R_i)$ is bounded by some   constant. We can identify $(r, b) \in X_1 \times X_2$ with the intersection point of $r$ and $b$.

Considering the product set system $\Sigma_1 \otimes \Sigma_2 = (X_1 \times X_2, \T)$, it is easy to see that for any convex set $C$, the set of  intersection  points between
lines of $X_1$ and $X_2$ that lie within $C$ is an element of $\T$.
The VC-dimension of $\Sigma_1 \otimes \Sigma_2$ is in fact unbounded.
Indeed, notice that any matching $\big\{(r_1,b_1), (r_2,b_2), \ldots, (r_k, b_k) \big\} \subset X_1 \times X_2$ is shattered by $\Sigma_1 \otimes \Sigma_2$.
Nevertheless, it is possible to construct small $\epsilon$-approximations for this
set system:
% the product of two set systems with VC-dimension bounded by a constant. Generalizing some earlier work by Chazelle \cite{C93},  Br{\"o}nnimann, Chazelle, and Matou{\v{s}}ek \cite{BCM} showed:

%Suppose that $X_1$ consists of a finite set of ``red'' lines in the plane, and the family $\R_1$ contains every subset $X'  \subseteq X_1$ such that $X'$ is precisely the subset of lines intersected by some line segment. Let $X_2$ consist of a finite subset of ``blue'' lines, and $\S_2 = (X_2, \R_2)$ be the set system induced by intersection with line segments, as above. Assume that each red line intersects each blue line, and that no three lines intersect in a point.

\lemma{\cmrx\cite{C93,BCM}\label{lem:product-sampling}}
{For $i = 1,2$ and $0 \leq \epsilon_i \leq 1$, let $A_i$ be an $\epsilon_i$-approximation for the finite set system $\Sigma_i = (X_i, \R_i)$. Then $A_1 \times A_2$ is an $(\epsilon_1 + \epsilon_2)$-approximation for $\Sigma_1 \otimes \Sigma_2$.}

We can apply this general result on $\Sigma_1 \otimes \Sigma_2$ to estimate  $V(X_1 \times X_2, C)$---defined to be
the number of intersections between lines in $X_1$ and $X_2$ that are contained in a query convex set $C$---by $\frac{|V(A_1 \times A_2, C)| \cdot |X_1| \cdot |X_2|}{|A_1| \cdot |A_2|}$. Lemma~\ref{lem:product-sampling} implies that
the error of this estimate can be bounded by
\[ \Big| \frac{V(X_1 \times X_2, C)}{|X_1| \cdot |X_2|} - \frac{V(A_1 \times A_2, C)}{|A_1| \cdot |A_2|} \Big| \leq \epsilon_1 + \epsilon_2.\]
The notion of a product of set systems
and Lemma~\ref{lem:product-sampling} can be generalized to more than two set systems \cite{BCM, Cha00}.

\Bnn{SHAPE FITTING AND CORESETS}

\noindent
Consider the scenario where the shape family $\F$ contains, as its elements, all possible $k$-point subsets of $\Re^d$; that is, each $F \in \F$ is a subset of $\Re^d$ consisting of $k$ points. If the function $\dist(\cdot, \cdot)$ is the Euclidean distance, then the corresponding shape fitting problem $(\Re^d, \F, \dist)$ is the well-known $k$-median problem. If $\dist (\cdot, \cdot)$ is the square of the Euclidean distance, then the shape fitting problem is the $k$-means problem. If the shape family $\F$ contains as its elements all hyperplanes in $\Re^d$, and $\dist (\cdot, \cdot)$ is the Euclidean distance, then   the corresponding shape fitting problem asks
for a hyperplane that minimizes the sum of the Euclidean distances from points in the given instance $P \subset \Re^d$.
The shape fitting problem as defined is just one of many versions that have been considered. In another well-studied version, given an instance $P \subset \Re^d$, the goal is
to find a shape that minimizes $\max_{p \in P} \dist(p,F)$.

Given an instance $P$, and a parameter $0 < \epsilon < 1$, an $\epsilon$-coreset $(S,w)$ ``approximates'' $P$   with respect to every shape $F$ in the given family $\F$. Such an $\epsilon$-coreset can be used to find a shape that approximately minimizes $\sum_{p \in P} \dist(p,F)$: one simply finds a shape that minimizes $\sum_{q \in S} w(q) \cdot \dist(q,F)$. For this approach to be useful, the size of the coreset needs to be small as well
as efficiently computable. Building on a long sequence of works,
Feldman and Langberg~\cite{FL11} (see also Langberg and Schulman~\cite{LS10})  showed the
existence of a function $f: \Re \rightarrow \Re$ such
that an $\epsilon$-approximation for a carefully constructed
set system associated with the shape fitting problem $(\Re^d, \F, \dist)$ and instance $P$ yields an $f(\epsilon)$-coreset for the instance $P$. For many shape fitting problems, this method often yields coresets with size guarantees that are not too much worse
than   bounds via more specialized arguments. We refer the reader
to the survey~\cite{BLK17} for further details.

% \Bnn{OPEN PROBLEMS}

% \noindent Here we give a few of the problems from the
% multitude of interesting questions that remain open.
% \begin{enumerate}

% \item \tindex{} Tusnady's problem:
% Given a set $P$ of $n$ points in $\Re^d$, one can two-color
% $P$ such that the discrepancy with respect to axis-parallel
% boxes is $O(\log^{d/2} n)$.
% \end{enumerate}

%\pagebreak

\A{EPSILON-NETS}

\noindent While an $\eps$-approximation of a set system $(X, \R)$
aims to achieve equality in the \emph{proportion} of points picked
from each set, often only a weaker threshold property is needed.
A set $N \subseteq X$  is called an $\eps$-net for $\R$
if it has a non-empty intersection with each set of $\R$ of
cardinality at least $\eps |X|$.
For all natural geometric set systems,
trivial considerations imply that any such $N$
must have size $\Omega(\frac{1}{\eps})$: one can always arrange the elements of $X$
into disjoint $\lfloor \frac{1}{\eps} \rfloor$ groups, each with at least $\eps |X|$ elements, such
that the set consisting of the elements
in each group is induced by the given geometric family.
%For the primal set system induced by intervals in $\Re$, it is easy
%to show the existence of an $N$ of size $\lfloor 1/\eps \rfloor$; less
%obvious is the existence of $O(\frac{1}{\eps})$ sized nets
%for the set systems induced by half-spaces in $\Re^2$.
%Surprisingly, this is also true for set systems
%induced by half-spaces in $\Re^3$.
While $\eps$-nets form the basis of many algorithmic and combinatorial
tools in discrete and computational geometry,
here we  present only two   applications,
one combinatorial and one algorithmic.

 %\vspace{1pc}

\Bnn{GLOSSARY}

\begin{gllist}
\item {\tindex{epsilon-net@$\varepsilon$-net} \trmbitx $\eps$-Nets:}\quad
Given a finite set system $(X, \R)$ and a parameter $0 \leq \eps \leq 1$,
a set $N \subseteq X$ is an $\eps$-net for $\R$ if $N \cap R \neq \emptyset$
for all sets $R \in \R$ with $|R| \geq \eps |X|$.

\item {\tindex{epsilon-net@$\varepsilon$-net!weak}\trmbitx Weak $\eps$-nets:}\quad
Given a set $X$ of points in $\Re^d$ and family of objects $\R$,
a set $Q \subseteq \Re^d$
is a \emph{weak $\eps$-net} with respect to $\R$
if $Q \cap R \neq \emptyset$ for all $R \in \R$ containing
at least $\eps |X|$ points of $X$. Note that in contrast
to $\eps$-nets, we do not require $Q$ to be a subset
of $X$.

\item {\tindex{semialgebraic set}\trmbitx Semialgebraic sets:}\quad
Semialgebraic sets are
subsets of $\Re^d$ obtained by taking Boolean operations such as
unions, intersections, and complements of sets of the
form $\{ x \in \Re^{d} \mid g(x) \geq 0\}$, where $g$ is a $d$-variate
polynomial in $\mathbb{R} \left[ x_{1}, \ldots, x_{d}\right]$.

\item {\tindex{epsilon-net@$\varepsilon$-net!Mnets}\trmbitx $\epsilon$-Mnets:}\quad
Given a set system $(X, \R)$ and a parameter $0 \leq \eps \leq 1$,
a collection of sets  $\mathcal{M} = \{X_1, \ldots, X_t\}$ on $X$
is an $\eps$-Mnet
of size $t$ if  $|X_i| = \Theta(\eps |X|)$ for all $i$, and for any set
$R \in \R$ with $|R| \geq \eps |X|$, there exists an index
$j \in \{1, \ldots, t\}$ such that $X_j \subseteq R$.
\end{gllist}

\vspace{-1pc}
\Bnn{EPSILON-NETS FOR ABSTRACT SET SYSTEMS}

\noindent The systematic study of $\eps$-nets
started with the breakthrough result of Haussler and Welzl~\cite{HW87},
who first showed the existence of $\eps$-nets
whose size was a function of the parameter $\eps$ and the
VC-dimension.
A different framework, with somewhat similar ideas and consequences, was independently introduced by Clarkson~\cite{C87}.
The result of Haussler and Welzl was later improved upon and extended in several ways:
the precise dependency on $\vcdim(\R)$ was improved,
the probabilistic proof in~\cite{HW87} was de-randomized
to give a deterministic algorithm,
and finer probability estimates were derived
for randomized constructions of $\eps$-nets.

\theorem{\cmrx\cite{HW87,KPW92}\tindex{epsilon-net theorem@$\varepsilon$-net theorem}\label{thm:epsnet}}
{Let $(X, \R)$ be a  finite set system, such that
$\pi_{\R}(m) = O(m^d)$ for a
fixed constant $d$, and let $\eps>0$ be a sufficiently small parameter. Then there exists
an $\eps$-net for $\R$ of size $\big(1+o(1)\big)\frac{d}{\epsilon} \log \frac{1}{\epsilon}$.
Furthermore, a uniformly chosen random sample of $X$
of the above size is an $\eps$-net with   constant probability.}

An alternate proof, though with   worse constants,
follows immediately from   $\eps$-approximations: use
Theorem~\ref{thm:epsapprox-VC} to compute an $\frac{\eps}{2}$-approximation
$A$ for $(X, \R)$, where $|A|=O(\frac{d}{\eps^2})$.
Observe that an $\frac{\eps}{2}$-net
for $(A, \R|_A)$ is an $\eps$-net for $(X, \R)$,
as for each $R\in \R$ with $|R| \geq \eps |X|$,
we have $\big| \frac{|R|}{|X|} - \frac{|R\cap A|}{|A|} \big| \leq \frac{\eps}{2}$
and so $\frac{|R\cap A|}{|A|} \geq   \frac{\eps}{2}$.
Now a straightforward random sampling argument with union bound (or an iterative
greedy construction)
gives an $\frac{\eps}{2}$-net for $\R|_A$, of total
size $O\big( \frac{1}{\eps} \log |\R|_A| \big) = O\big(\frac{d}{\eps} \log \frac{d}{\eps} \big)$.

\theorem{\cmrx\cite{AS08}\label{thm:epsnets-VC}}
{Let $(X, \R)$ be a finite set system with ${\pi_{\R}(m) = O(m^d)}$ for a
constant $d$, and $0 < \epsilon, \gamma \leq 1$ be given parameters. Let $N \subseteq X$ be a  set of size
\[ \max \Big\{ \frac{4}{\eps} \log \frac{2}{\gamma}, \  \frac{8d }{\eps} \log \frac{8d}{\eps} \Big\} \]
%\todo{this is   true, \\but not stated anywhere in literature. who to cite it to??}
chosen uniformly at random. Then $N$ is an $\epsilon$-net
 with probability at least $1 -\gamma$.}

\theorem{\cmrx\cite{BCM}}
{Let $(X, \R)$ be a finite set system such that ${\vcdim(\R) = d}$,
and $\eps> 0$ a given parameter.
%Assume that given any $Y \subseteq X$, all the sets
Assume that for any $Y \subseteq X$, all sets
in $\R|_Y$ can be computed explicitly in time $O\big(|Y|^{d+1}\big)$.
Then an $\eps$-net of size $O\big( \frac{d}{\eps} \log  \frac{d}{\eps}  \big)$
can be computed deterministically in time $O(d^{3d}) \cdot (\frac{1}{\eps} \log \frac{1}{\eps})^d \cdot |X|$.}

It was shown in~\cite{KPW92} that for any $0 < \eps \leq 1$, there exist $\eps$-nets
of size $\max \big\{2, \lceil \frac{1}{\eps} \rceil -1 \big\}$ for any set system $(X, \R)$
with $\vcdim(\R)=1$.
For the case when $\vcdim(\R) \geq 2$, the quantitative bounds of Theorem~\ref{thm:epsnet} are near-optimal, as the following
construction shows.
For a given integer $d \geq 2$ and a real $\eps>0$,
set $n = \Theta \big(\frac{1}{\eps} \log \frac{1}{\eps} \big)$
and construct a random $\eps n$-uniform set system by choosing
$\Theta\big( \frac{1}{\eps^{d+\gamma-1}} \big)$ sets uniformly from all possible
sets of size $\eps n$,
%each set uniformly with probability $\frac{ \eps^{1-d-\gamma} }{{n \choose \eps n}}$,
where $\gamma$ is  sufficiently small.
It can be shown that, with constant probability, this set system has VC-dimension
at most $d$
and any $\eps$-net for it must have large size.

\theorem{\cmrx\cite{KPW92}\label{thm:vcdimlowerbound}}
{Given any $\eps>0$ and integer $d \geq 2$,
%\footnote{This is not merely a technical
%condition; it is easy to see the existence of $O(\frac{1}{\eps})$ sized
%nets for set systems with VC-dimension $1$.}
there exists a set system $(X, \R)$ such
that $\vcdim(\R) \leq d$ and any $\eps$-net for $\R$
has size at least $\big( 1-\frac{2}{d}+\frac{1}{d(d+2)}+o(1) \big) \frac{d}{\eps} \log \frac{1}{\eps}$.}

\noindent Over the years it was realized that the shatter function of a set system
is too crude a characterization
for purposes of $\eps$-nets, and that the existence
of smaller sized $\eps$-nets can be shown if one further knows the distribution
of sets of any fixed size in the set system.
This was first understood for the case of geometric dual
set systems in $\Re^2$ using spatial partitioning techniques, initially
in the work of Clarkson and Varadarajan~\cite{CV07} and then in its improvements
by Aronov~\emph{et al.}~\cite{AES10}.
Later it was realized by Varadarajan~\cite{V09,V10}
and in its improvement by Chan \emph{et al.}~\cite{CGKS12} that one could
avoid spatial partitioning altogether, and get improved
bounds on sizes of $\eps$-nets in terms of the shallow-cell
complexity of a set system.

\theorem{\cmrx\cite{V10, CGKS12}\label{theorem:shallowcellcomplexityepsnets}}
%\dtodo{problem: papers only cite the version with n, instead of 1/eps}
{Let $(X, \R)$ be a set system with shallow-cell complexity
$\varphi_{\R}(\cdot)$, where $\varphi_{\R}(n) = O(n^d)$ for
some constant $d$. Let $\eps > 0$ be a given parameter.
Then there exists an $\eps$-net\footnote{The bound in these papers is stated as
$O\big(\frac{1}{\eps} \log \varphi_{\R}(|X|) \big)$, which does not require
the assumption that $\varphi_{\R} (n)= O(n^d)$ for some constant $d$.  However,
standard techniques using $\eps$-approximations
imply the stated bound; see~\cite{V09,KMP16} for details.}
for $\R$ of size $O\big(\frac{1}{\eps} \log \varphi_{\R}(\frac{1}{\eps})\big)$.
Furthermore, such an $\eps$-net can be computed in deterministic
polynomial time.}

We sketch a simple proof of the above theorem
due to Mustafa \emph{et al.}~\cite{MDG16}. For simplicity,
assume that $|R| = \Theta(\eps n)$ for all $R \in \R$.
Let $\P \subseteq \R$ be a maximal $\frac{\eps n}{2}$-separated
system, of size $|\P| = O\big( \frac{1}{\eps} \varphi_{\R}(\frac{1}{\eps})\big)$
by Lemma~\ref{lem:shallowpacking}. By the maximality
of $\P$, for each $R \in \R$  there exists a $P_R \in \P$ such that
$|R \cap P_R| \geq \frac{\eps n}{2}$, and thus a
set $N$ which is a $\frac{1}{2}$-net for each
of the $|\P|$ set systems $(P, \R|_P)$, $P \in \P$, is an $\eps$-net for $\R$.
Construct the set $N$ by picking each point of $X$ uniformly with
probability $\Theta \big(\frac{1}{\eps n} \log \varphi_{\R} (\frac{1}{\eps}) \big)$.
For each $P \in \P$, $P \cap N$ is essentially a random subset
of size $\Theta \big(\log \varphi_{\R} (\frac{1}{\eps}) \big)$,
and so by Theorem~\ref{thm:epsnets-VC}, $N$
fails to be a $\frac{1}{2}$-net for $\R|_P$ with probability
$O\big(\frac{1}{\varphi_{\R}(1/\eps)}\big)$.
By linearity of expectation, $N$ is a $\frac{1}{2}$-net
for all but expected $O\big(\frac{1}{\varphi_{\R}(1/\eps)}\big) \cdot |\P| = O(\frac{1}{\eps})$ sets of $\P$, and for those a $O(1)$-size
$\frac{1}{2}$-net can be constructed individually (again by Theorem~\ref{thm:epsnets-VC})
and added to $N$, resulting
in an $\eps$-net of expected size $\Theta \big(\frac{1}{\eps} \log \varphi_{\R} (\frac{1}{\eps}) \big)$.
%In particular,
%$N_1$ fails to be a $\frac{1}{2}$-net for
%
%Let $N_1$ be a subset of size
%$\Theta \big(\frac{1}{\eps} \log \varphi_{\R} (\frac{1}{\eps}) \big)$
%chosen uniformly at random from $X$.
% implies, with $\gamma = $
%and $\eps' = \frac12$,
%By  and union bound, ,

Furthermore, this bound can be shown to be near-optimal by
generalizing the random construction  used in Theorem~\ref{thm:vcdimlowerbound}.

\theorem{\cmrx\cite{KMP16}}
{Let $d$ be a fixed positive integer and let ${\varphi:\Ne \to \Re^{+}}$ be any
submultiplicative function\footnote{A function
$\varphi: \Re^+ \to \Re^+$ is called {\em submultiplicative} if
$(a)$ $\varphi^{\alpha}(n)\le \varphi(n^{\alpha})$ for any $0<\alpha<1$
and a sufficiently large positive $n$, and
$(b)$ $\varphi(x)\varphi(y)\ge \varphi(xy)$
for any sufficiently large $x,y\in \Re^+$.} with $\varphi(n) = O(n^d)$
for some constant $d$.
Then, for any $\eps>0$, there exists a set system  $(X,\R)$ with
shallow-cell complexity $\varphi(\cdot)$, and for which
any $\eps$-net has size
$\Omega\big(\frac{1}{\epsilon} \log \varphi(\frac 1{\epsilon})\big)$.}

On the other hand, there are examples of natural set systems
with high shallow-cell complexity and yet with small $\eps$-nets~\cite{Mat16}:
for a planar undirected graph $G = (V, E)$, let $\R$ be the set system
on $V$ induced by shortest paths in $G$; i.e.,
for every pair of vertices $v_i, v_j \in V$, the set $R_{i,j} \in \R$ consists
of the set of vertices on the shortest path between $v_i$ and $v_j$. Further,
assume that these shortest paths are unique for every pair of vertices.
Then $(V, \R)$ has $\eps$-nets of size $O\big( \frac{1}{\eps} \big)$~\cite{KPR93},
and yet $\varphi_{\R}(n) = \Omega(n)$
can be seen, e.g., by considering the star graph.
As we will see in the next part,
the primal set system induced by axis-parallel rectangles
is another example with high
shallow-cell complexity and yet small  $\eps$-nets.

 The proof in~\cite{V10,CGKS12} presents a randomized
method to construct an $\eps$-net $N$ such that
each element $x \in X$ belongs to $N$ with
probability $O\big(\frac{1}{\eps |X|} \log \varphi_{\R}(\frac{1}{\eps})\big)$. This
implies the following more general result.

\corollary{\cmrx\cite{V10,CGKS12}}
{Let $(X, \R)$ be a set system with shallow-cell complexity
$\varphi_{\R}(\cdot)$, and $\eps > 0$ be
a given parameter. Further let $w: X \rightarrow \Re^+$ be weights
on the elements of $X$, with $W = \sum_{x \in X} w(x)$.
Then there exists an $\eps$-net
for $\R$ of total weight $O\big(\frac{W}{\eps |X|} \log \varphi_{\R}(\frac{1}{\eps}) \big)$.}

The notion of $\eps$-Mnets of a
set system $(X, \R)$, first
defined explicitly and studied in Mustafa and Ray~\cite{MR14}, is related to both
$\eps$-nets (any
transversal of the sets in an $\eps$-Mnet is an $\eps$-net for $ \R $)
as well as  the so-called \emph{Macbeath regions} in convex
geometry (we refer the reader to the surveys~\cite{BL88, Bar07} for more details on Macbeath regions,
and to Mount \etal~\cite{AFM17} for some recent applications). The following theorem concerns $\eps$-Mnets
with respect to volume  for the primal set system
induced by half-spaces.
 
%\pagebreak

\theorem{\cmrx\cite{BCP93}}
{Given a compact convex body $K$ in $\Re^d$ and
a parameter $0 < \eps < \frac{1}{(2d)^{2d}}$, let
$\R$ be the primal set system on $K$ induced by half-spaces in $\Re^d$,
equipped with Lebesgue measure.
There exists an $\eps$-Mnet for $\R$ of
size $O \big( \frac{1}{\eps^{1-\frac{2}{d+1}}} \big)$. Furthermore,
the sets in the $\eps$-Mnet are pairwise-disjoint convex bodies lying
in $K$.}

The role of shallow-cell complexity carries over
to the bounds on $\eps$-Mnets; the proof of the following theorem
uses the packing lemma (Lemma~\ref{lem:shallowpacking}).

\theorem{\cmrx\cite{DGJM16} \label{theorem:shallowcellcomplexityMnets}}
{Given a set $X$ of points in $\Re^d$, let
$\R$ be the primal set system on $X$ induced
by a family of semialgebraic
sets in $\Re^{d}$ with shallow-cell complexity $\varphi_{\R}(\cdot)$,
where $\varphi_{\R}(n) = O(n^t)$ for some constant $t$. Let $\eps > 0$
be a given parameter.
Then there exists an $\eps$-Mnet for $\R$
of size $O\big(\frac{1}{\eps} \varphi_{\R} ( \frac{1}{\eps})\big)$,
where the constants in the asymptotic notation depend on  the degree and
number of inequalities defining the semialgebraic sets.}

Together with bounds on shallow-cell complexity
for half-spaces (Lemma~\ref{lemma:shallowcellforhalfspaces}), this
implies the existence of $\eps$-Mnets
of size $O \big( \frac{1}{\eps^{\lfloor d/2 \rfloor}} \big)$
for the primal set system induced by half-spaces
on a finite set of points in $\Re^d$.
Further, as observed in~\cite{DGJM16}, Theorem~\ref{theorem:shallowcellcomplexityMnets}
implies Theorem~\ref{theorem:shallowcellcomplexityepsnets} for semialgebraic
set systems by a straightforward use of random sampling and the union bound.

\Bnn{EPSILON-NETS FOR  GEOMETRIC SET SYSTEMS}

\noindent
 We now turn to set systems, both primal and dual, induced
by geometric objects in $\Re^d$. 
The existence of $\eps$-nets of size $O(\frac{1}{\eps} \log \frac{1}{\eps})$ for several geometric set systems follow from the early breakthroughs of Clarkson~\cite{C87}
and Clarkson and Shor~\cite{CS89} via the use of random sampling together
with spatial partitioning.
For the case of primal and dual set systems, it turns out that all known asymptotic bounds 
on sizes of $\eps$-nets  
follow from Theorem~\ref{theorem:shallowcellcomplexityepsnets} and
bounds on shallow-cell complexity (Table~\ref{table:shallowcellcomplexity}).
The relevance of shallow-cell complexity for $\eps$-nets was realized
after considerable effort was spent on inventing
a variety of specialized techniques for constructing
 $\eps$-nets for geometric set systems.
These techniques and ideas
have their own advantages, often yielding algorithms with low running times
and low constants hidden
in the asymptotic notation. Table~\ref{table:geometricepsnets}
lists the most precise upper bounds known for many natural geometric
set systems; all except one are, asymptotically, direct consequences
of Theorem~\ref{theorem:shallowcellcomplexityepsnets}.
The exception is the case of the primal set system
induced by the family $\R$ of axis-parallel rectangles in the plane,
which have shallow-cell
complexity   $\varphi_{\R}(n) = n$, as for any integer $n$  there exist
a set $X$ of $n$ points in $\Re^2$ such that the number
of subsets of $X$ of size at most two induced by $\R$ is $\Theta(n^2)$.
However, Aronov \emph{et al.}~\cite{AES10} showed that there exists
another family of objects\footnote{Constructed as follows: let $l$ be a  vertical line that divides $X$
into two equal-sized subsets, say $X_1$ and $X_2$;
then add to $\R'$ all subsets of $X$ induced
by axis-parallel rectangles with one
 vertical boundary edges lying on $l$. Add recursively
subsets to $\R'$ for $X_1$ and $X_2$.} $\R'$ with $\varphi_{\R'}(n) = O(\log n)$, such
that an $\frac{\eps}{2}$-net for the primal set system on $X$ induced by $\R'$ is an $\eps$-net
for the one induced by $\R$; now $\eps$-nets
of size $O \big(\frac{1}{\eps} \log \log \frac{1}{\eps} \big)$  for the primal set system induced by $\R$ follow by applying Theorem~\ref{theorem:shallowcellcomplexityepsnets}
on $\R'$.

Precise sizes of $\eps$-nets for some constant values
of $\eps$ have been studied for the primal set system induced by
axis-parallel rectangles and disks in $\Re^2$~\cite{AAG14}.
It is also known that the visibility
set system for a simple polygon $P$
and a finite set of guards $G$---consisting of all sets $S_p$, where
$S_p  $ is the set of points of $G$ visible from
$p \in P$---admits $\eps$-nets of size $O\big(\frac{1}{\eps} \log \log \frac{1}{\eps}\big)$~\cite{KK11}.
In the case where the underlying base set
is $\Re^d$, bounds better than those following
from Theorem~\ref{theorem:shallowcellcomplexityepsnets}
are known from the theory of geometric coverings.

\theorem{\cmrx\cite{R57}}
{Let $K \subset \Re^d$ be a bounded convex body, 
and let $Q = [-r, r]^d$ be a cube of side-length $2r$, where $r \in \Re^+$.  
Let $\R$ be the primal set system induced by translates
of $K$ completely contained in $Q$. 
Then there exists
a hitting set $P \subset Q$  for $\R$ 
of size at most
$$\frac{r^d}{\vol(K)} \cdot \left( d \ln d + d \ln \ln d + 5d \right).$$
}

Note that Theorem~\ref{theorem:shallowcellcomplexityepsnets}
cannot  be used here, as translates of a convex set have unbounded VC-dimension and exponential shallow-cell complexity. 
Furthermore, even for the  
case where $K$ is a unit ball in $\Re^d$, Theorem~\ref{theorem:shallowcellcomplexityepsnets} 
would give a worse bound of $O\left(\frac{r^d}{\vol(K)} \cdot d^2 \log r\right)$.

%\vspace{-0.5pc}

%\pagebreak

\begin{table}[h]
\viiipt
\baselineskip=11pt
\renewcommand{\arraystretch}{1.0}
\Table{Sizes of $\eps$-nets for both primal (P) and dual (D) set   systems (ceilings/floors and lower-order terms are omitted for clarity).}
{\begin{tabular}{| l | l | l | l |}
    \hline
\rule[-4pt]{0pt}{13pt}{\hvbviii Objects}
    & {\hvbviii SETS}
    & {\hvbviii UPPER BOUND}
    & {\hvbviii LOWER BOUND}
    \\ \hline
\rule[0pt]{0pt}{9pt}Intervals
\label{table:geometricepsnets}
    & P/D
    & $\frac{1}{\eps}$
    & $\frac{1}{\eps}$
    \\
Lines, $\Re^2$
    & P/D
    & $\frac{2}{\eps} \log \frac{1}{\eps}$ \hfill\cite{HW87}
    & $\frac{1}{2\eps} \frac{\log^{1/3} \frac{1}{\eps}}{\log \log \frac{1}{\eps}}  $ \hfill\cite{BS17}
    \\
Half-spaces, $\Re^2$
    & P/D
    & $\frac{2}{\eps} - 1$ \hfill \cite{KPW92}
    & $\frac{2}{\eps} - 2$ \hfill \cite{KPW92}
    \\
Half-spaces, $\Re^3$
    & P/D
    & $O(\frac{1}{\eps})$ \hfill \cite{MSW90}
    & $\Omega(\frac{1}{\eps})$ \hfill
    \\
Half-spaces, $\Re^d$, $d \geq 4$
    & P/D
    & $ \frac{d}{\eps} \log \frac{1}{\eps}$ \hfill \cite{KPW92}
    & $\frac{\lfloor d/2 \rfloor - 1}{9}\frac{1}{\eps} \log \frac{1}{\eps}$ \hfill $\substack{\cite{PT13}\\
    								\ \ \ \cite{KMP16}}$
    \\
Disks, $\Re^2$
    & P
    & $\frac{13.4}{\eps}$ \hfill \cite{BGMR16}
    & $\frac{2}{\eps}-2$ \hfill \cite{KPW92}
    \\
Balls, $\Re^3$
    & P
    & $\frac{2}{\eps} \log \frac{1}{\eps}$
    & $\Omega( \frac{1}{\eps})$
    \\
Balls, $\Re^d$, $d \geq 4$
    & P
    & $\frac{d+1}{\eps} \log \frac{1}{\eps}$ \hfill \cite{KPW92}
    & $\frac{\lfloor d/2 \rfloor - 1}{9}\frac{1}{\eps} \log \frac{1}{\eps}$ \hfill \cite{KMP16}
    \\
Pseudo-disks, $\Re^2$
    & P/D
    & $O(\frac{1}{\eps})$ \hfill \cite{PR08}
    & $\Omega(\frac{1}{\eps})$
    \\
Fat triangles, $\Re^2$
    & D
    & $O(\frac{1}{\eps} \log \log^* \frac{1}{\eps})$ \hfill \cite{AES10}
    & $\Omega(\frac{1}{\eps})$
    \\
Axis-par. rect., $\Re^2$
    & D
    & $\frac{5}{\eps} \log \frac{1}{\eps}$ \hfill \cite{HW87}
    & $\frac{1}{9}\frac{1}{\eps} \log \frac{1}{\eps}$ \hfill \cite{PT13}
    \\
Axis-par. rect., $\Re^2$
    & P
    & $O(\frac{1}{\eps} \log \log \frac{1}{\eps})$ \hfill \cite{AES10}
    & $\frac{1}{16} \frac{1}{\eps} \log \log \frac{1}{\eps}$ \hfill \cite{PT13}
    \\
Union $\k_{\R}(\cdot)$, $\Re^2$
    & D
    & $O\big(\frac{\log ( \eps \cdot \k_{\R}( 1/\eps) )}{\eps}\big)$ \hfill \cite{AES10}
    & $\Omega(\frac{1}{\eps})$
    \\
Convex sets, $\Re^d$, $d \geq 2$
	& P
	& $|X| - \eps |X|$
	& $|X| - \eps |X|$
	\\
    \hline
\end{tabular}}
\renewcommand{\arraystretch}{1}
\xpt
\baselineskip=12pt
\end{table}

%\footnotetext{Here $w(s)$ is the minimum number $k$
%so that $k^{A_k(2)} > s$, where $A_k$ is the $k$-th
%function in the Ackermann hierarchy.}

Lower bounds for $\epsilon$-nets for geometric
set systems are implied by the following connection, first observed by Alon~\cite{A12},
between $\eps$-nets and density version of statements
in Ramsey theory. Given a function $f: \Ne^+ \to \Ne^+$,
let $(X, \R)$, $|X|=n$, be a set system
with the Ramsey-theoretic property that for any
$X' \subset X$ of size $\frac{n}{2}$, there exists
a set $R \in \R$ such that $|R| \geq f(n)$
and $R \subseteq X'$. Then any $\frac{f(n)}{n}$-net $N$ for
$(X, \R)$ must have size at least $\frac{n}{2}$, as
otherwise the set $X \setminus N$ of size at least $\frac{n}{2}$
would violate the Ramsey property. As $\frac{n}{2} = \omega\big( \frac{n}{f(n)}\big)$
for any monotonically increasing function $f(\cdot)$ with
$f(n) \to \infty$ as $n \to \infty$, this gives
a super-linear lower bound on the size of any $\frac{f(n)}{n}$-net;
the precise lower bound will depend on the function $f(\cdot)$.
%;e.g.,   $f(n) = \Omega( \log n)$ gives the   bound
%of $\Omega\big(\frac{1}{\eps} \log \frac{1}{\eps}\big)$.
Using this relation, Alon~\cite{A12} showed a super-linear lower bound
for $\eps$-nets for the primal set system induced by lines,
for which the corresponding Ramsey-theoretic statement
is the density version of the Hales-Jewett theorem.
By Veronese maps\footnote{Map each point $p: (p_x, p_y) \in \Re^2$
to the point $f(p) = (p_x, p_y, p_xp_y, p_x^2, p_y^2) \in \Re^5$,
and each line $l: ax+by=c$ to the half-space
$f(l): (-2ac)\cdot x_1 + (-2bc) \cdot x_2 + (2ab) \cdot x_3 + a^2 \cdot
x_4 + b^2 \cdot x_5 \leq -c^2$.
Then it can be verified by a simple calculation that
a point $p \in \Re^2$ lies on a line  $l$ if and only
if the point $f(p) \in \Re^5$ lies in the half-space $f(l)$.}, this   implies a nonlinear bound
for $\eps$-nets for the primal set system induced by
half-spaces in $\Re^5$. Next,
%overturning conventional wisdom on the subject,
Pach and Tardos~\cite{PT13}
showed that, for any $\eps > 0$ and large enough integer $n$,
there exists a set $X$ of $n$ points in $\Re^4$ such that
any $\eps$-net for the primal set system on $X$ induced
by half-spaces must have size  at least $\frac{1}{9\eps} \log\frac{1}{\eps}$;
when $\frac{1}{\eps}$ is a power of two, then it improves
to the lower bound of $\frac{1}{8\eps} \log\frac{1}{\eps}$.
 See  Table~\ref{table:geometricepsnets} for all known lower bounds.

\medskip\noindent{\trmbitx Weak $\eps$-nets.}
When the net for a given
primal geometric set system $(X, \R)$  need not be a subset of $X$---i.e., the case of weak $\eps$-nets---one can sometimes get smaller bounds.
For example, $O(\frac{1}{\eps})$ size  weak $\eps$-nets
exist for the primal set system induced by balls in $\Re^d$~\cite{MSW90}.
We outline a different construction than the one in~\cite{MSW90},
as follows.
Let  $B$ be the smallest radius ball containing
a set $X'$ of at least $\eps |X|$ points of $X$
and no point of the current weak $\eps$-net $Q$ (initially $Q = \emptyset$). Now add a set $Q' \subseteq \Re^d$
of $O(1)$ points   to $Q$ such
that any ball, of radius at least that of $B$,
intersecting $B$ must contain a point of $Q'$, and
compute a weak $\eps$-net for $X \setminus X'$.
Weak $\eps$-nets of size $O \big( \frac{1}{\eps} \log \log \frac{1}{\eps} \big)$ exist
for the primal set system induced by axis-parallel rectangles in $\Re^d$,
for $d \geq 4$~\cite{E10}.

The main open question at this time on weak $\eps$-nets is for the primal set system
induced on a set $X$ of $n$ points by the family $\C$
of all convex objects in $\Re^d$. Note that if $X$ is in convex position, then
any $\eps$-net for this   set system  must have size at least $(1-\eps) n$.
All currently known upper bounds depend exponentially
on the dimension $d$.
In Alon \etal~\cite{ABFK92}, a
bound of $O\big( \frac{1}{\eps^2} \big)$ was shown for this problem
for $d=2$ and $O(\frac{1}{\eps^{d+1}})$ for $d \geq 3$.
This was improved by Chazelle \etal~\cite{CEGGSW93},
and then slightly further via an elegant proof
by Matou{\v{s}}ek and Wagner~\cite{MW02}.

\theorem{\cmrx\cite{MW02}\label{thm:weakepsnets}}
{Let $X$ be a finite set of points in $\Re^d$, and let $0 < \eps \leq 1$
be a given parameter. Then there exists a weak $\eps$-net
for the primal set system induced by convex objects   of size $O\big( \frac{1}{\eps^d} \log^{a} (\frac{1}{\eps}) \big)$, where $a = \Theta \big(d^2 \ln (d+1) \big)$.
Furthermore, such a net can be computed in time
$O\big(n \log \frac{1}{\eps}  \big) $.}
The above theorem---indeed many of the weak $\eps$-net
constructions---are based on the following two ideas.
First, for a parameter $t$ that is chosen carefully,
construct a partition $\P = \{ X_1, \ldots, X_t \}$ of $X$
such that $(a)$ $|X_i| \leq \lceil \frac{n}{t} \rceil$ for all $i$,
and $(b)$ for any integer $k \geq 1$, there exists a point set $Q_k$
of small size
such that any convex object having non-empty
intersection with at least $\eps k$
sets of $\P$ must contain a point of $Q_k$.
Note that   $Q_t$ is   a weak $\eps$-net, as
any convex set  containing $\eps n$ points must intersect
at least $\frac{\eps n}{(n/t)} = \eps t $ sets.
%A convex set intersecting more $X_i$'s is easier to hit,
%and therefore the size of $Q_k$
%is a decreasing function of $k$.
Second, compute recursively
a weak $\eps'$-net $Q'_i$ for each $X_i$, for a suitably determined value of $\eps'$.
If a convex set $C$ is not hit by $\bigcup Q'_i$,
it contains at most $\frac{\eps' n}{t}$ points from each
set of $\P$, and so has non-empty intersection with
 at least $\frac{\eps n}{(\eps' n/t)} = \frac{t \eps}{\eps'}$ sets of $\P$.
Then $\bigcup Q'_i$ together with $Q_{\frac{t}{\eps'}}$ is a weak $\eps$-net;
fixing the trade-off
parameters $t, \eps'$ gives the final bound.
Theorem~\ref{thm:weakepsnets} uses simplicial partitions
for $\P$, and   centerpoints of  some representative
points from each set of $\P$ as the set $Q_k$.

There is a wide gap between the best known upper and lower bounds.
Matou{\v{s}}ek~\cite{M02} showed the existence
of a set $X$ of points in $\Re^d$ such that any weak $\frac{1}{50}$-net
for the set system induced by convex objects on $X$ has size
$\Omega\big(e^{\frac{\sqrt{d}}{2}}\big)$. For arbitrary
values of $\eps$, the current best lower bound is the following.

\vspace{-0.2pc}

\theorem{\cmrx\cite{BMV11}}
{For every $d \geq 2$ and every $\eps > 0$, there exists a set
$X$ of points in $\Re^d$ such that any weak $\eps$-net
for the primal set system induced on
$X$ by convex objects has size $\Omega\big(\frac{1}{\eps} \log^{d-1}\frac{1}{\eps}\big)$.}

There is a relation between weak $\eps$-nets induced by convex
sets and $\eps$-nets for the primal set system induced by intersections of half-spaces,
though the resulting size of the weak $\eps$-net
is still exponential in the dimension~\cite{MR08}.
The weak $\eps$-net problem is closely related
to an old (and still open) problem of Danzer and Rogers, which
asks for the area of the largest convex region avoiding a given set of $n$ points in
a unit square (see~\cite{PT12} for  a history of the problem).
Better bounds for weak $\eps$-nets for primal
set systems induced by convex  objects are known for special cases: an upper bound
of $O\big( \frac{1}{\eps} \alpha (\frac{1}{\eps}) \big)$ when $X$ is a set of
points in $\Re^2$ in convex position~\cite{AKNSS08}; optimal bounds when $\eps$ is a large constant~\cite{MR09b}; a bound
of $O\big(\frac{1}{\eps} (\log \frac{1}{\eps})^{\Theta(d^2 \ln d)}  \big)$ when the points lie on a moment curve in $\Re^d$~\cite{MW02}.

\vspace{-0.5pc}

\A{APPLICATIONS OF EPSILON-NETS}
\noindent
As $\eps$-nets capture some properties of random
samples with respect to a set system, a natural use of $\eps$-nets
has been for derandomization; the
best deterministic combinatorial algorithms  for linear programming~\cite{CM96, C16}
are  derived via derandomization using $\eps$-nets.
Another   thematic use originates
from the fact that an $\eps$-net   of a set system $(X, \R)$
can be viewed as a hitting set for sets in $\R$
of size at least $\eps |X|$, and so is relevant for
many types of covering optimization problems; a recent
example is the beautiful work of Arya \etal~\cite{AFM12} in approximating
a convex body by a polytope with few vertices.
At first glance,  the
restriction that an $\eps$-net   only guarantees to hit  sets of size at least $\eps |X|$
narrows its applicability. A breakthrough idea, with countless applications,
has been to first assign \emph{multiplicities} (or \emph{weights})
to the elements of $X$ such that all multisets have large size;
then $\eps$-nets can be used to ``round'' this to get a solution.
Lastly, $\eps$-nets can be used for constructing
spatial partitions that enable the use of divide-and-conquer
methods; indeed, one of the earliest applications introducing
 $\eps$-nets was by Clarkson~\cite{C88} to construct a spatial partitioning data-structure
for answering nearest-neighbor queries.
%in algorithmic and combinatorial applications.

%\Bnn{SPATIAL PARTITIONING}
\Bnnnr{SPATIAL PARTITIONING}
\noindent
Consider the set system $(\H, \R)$
where the base set $\H$ is a set of $n$ hyperplanes in $\Re^d$,
and $\R$ is the  set system induced by intersection
of simplices in $\Re^d$ with $\H$. An $\eps$-net for $\R$ consists of a subset $\H'$
such that any simplex intersecting at least $\eps n$ hyperplanes
of $\H$ intersects a hyperplane in $\H'$.
This implies that for any simplex $\Delta$
lying   in the interior of a cell
in the arrangement of $\H'$, the number
of hyperplanes of $\H$ intersecting $\Delta$ is
less than $\eps n$. One can further partition
each cell in the arrangement of $\H'$ into simplices,
leading to the powerful
concept of \emph{cuttings}. After a series of papers in the
1980s and early 1990s~\cite{CF90, mat91b},
the following is the best  result in terms
of both   combinatorial and algorithmic bounds.

\vspace{-0.04in}

\theorem{\cmrx\cite{C93}\tindex{cutting}\label{thm:cuttings}}
{Let $\H$ be a set of $n$ hyperplanes in $\Re^d$, and $r\geq1$
a given parameter. Then there exists
a partition of $\Re^d$ into  $O( r^d )$ interior-disjoint simplices,
such that the interior of each simplex intersects at most $\frac{n}{r}$
hyperplanes of $\H$. These  simplices,  together
with the list of hyperplanes intersecting the interior
of each simplex, can be found deterministically in time $O  ( nr^{d-1} )$.}

%\noindent
%Such a partition is called a $\frac{1}{r}$-cutting of $\H$,
%and the size of a cutting is the number of simplices in it.
There are many extensions of such a partition, called a
$\frac{1}{r}$-cutting, known for objects
other than hyperplanes; see Chapter~\chapArrg. Here we state just one such result.

 %The size can be shown to depend on the complexity of the arrangement induced by $\H$.

%\begin{theorem}[\cite{CF90}]
%Let $\H$ be a set of $n$ hyperplanes in $\Re^d$, and $\Delta$ a simplex.
%Then there exists a $\frac{1}{r}$-cutting inside $\Delta$, of
%size $O\big(T (\frac{r}{n}))^d + r^{d-1}\big)$, where $T$ is the number
%of vertices induced by $\H$ inside $\Delta$.
%\end{theorem}
%
%\begin{theorem}[\cite{BS95}]
%Let $\S$ be a set of $n$ line segments in $\Re^2$,
%and let $m$ denote the number of intersecting pairs of segments in $\S$.
%Then, for   any $r \leq n$, there is a $\frac{1}{r}$-cutting of $\S$ of size at most $O \big(r + \frac{mr^2}{n^2} \big)$.
%\label{thm:cuttingsimplices2d}
%\end{theorem}

\theorem{\cmrx\cite{BS95, P97}\tindex{cutting!for simplices}\label{thm:cuttingsimplices}}
{Let $\S$ be a set of $n$ $(d-1)$-dimensional simplices in $\Re^d$
and let $m = m(\S)$ denote the number of $d$-tuples of $\S$ having a point in common.
Then, for any $\eps>0$ and any given
parameter $r \geq 1$, there exists a $\frac{1}{r}$-cutting of $\S$ with
the number of simplices at most
 $\displaystyle O\left(r + \frac{mr^2}{n^2}\right)$
for $d=2$, and   $\displaystyle O\left(r^{d-1+\eps} + \frac{mr^d}{n^d}\right)$
for $d \geq 3$.}

Cuttings have found  countless applications, both combinatorial and algorithmic,
for their role in divide-and-conquer arguments. A paradigmatic combinatorial
use for upper-bounding purposes, initiated in a seminal
paper by Clarkson \etal~\cite{CEGSW90}, is   using cuttings
to partition $\Re^d$ into simplices, each of which forms an independent
sub-problem where one can apply a worse---and often purely combinatorial---bound.
The sum of this bound over all simplices together with accounting
for interaction  on the boundaries of the simplices gives an upper bound.
This remains  a key technique  for bounding incidences
between points and various geometric objects (see the book~\cite{G16}),
as well as for many
Tur\'an-type problems on geometric configurations (see~\cite{MP16} for a recent example).
%A first application of this idea~\cite{CEGSW90}, devised for an elegant
%proof of the Szmer\'edi-Trotter incidence theorem -- that any set
%  $P$ of $n$ points and $L$ of $m$ lines
%have $O\big(n^{\frac{2}{3}}m^{\frac{2}{3}} + n + m\big)$
%incidences between them -- is the following: for a suitable value of $r$,
%construct a $\frac{1}{r}$-cutting $\Xi$ for $L$; then the total number of incidences
%between $P$ and $L$ can be bounded recursively within each $\Delta \in \Xi$ to
%give the required overall bound (one also has to account for the incidences
%occuring on the boundary of the simplices of $\Xi$). A similar use
%gives near-optimal bounds for the
%Zarankiewicz problem for intersection hypergraphs~\cite{MP16}.
Algorithmically, cuttings have proven invaluable for divide-and-conquer
based methods for point location, convex hulls, Voronoi diagrams, combinatorial
optimization problems, clustering,
range reporting and range searching.
An early use was for the half-space range searching
problem, which asks for pre-processing a finite set $X$ of points in $\Re^d$
such that one can efficiently count the set of points
of $X$ contained in any query half-space~\cite{M93}.
The current best data structure~\cite{AC09}
for the related problem of reporting points contained in a query half-space
is also based on cuttings; see Chapter~\chapRange.
%, and can answer half-space range reporting
%queries using $O(n)$ space and time $O(\log n + k)$, where $k$ is the number of reported
%points, and with $O(n \log n)$ expected pre-processing time.

Finally, we state one consequence of a beautiful
result of Guth~\cite{G15} which achieves spatial partitioning
for more general objects, with a topological
approach replacing the use of $\epsilon$-nets: given a set $\H$ of $n$ $k$-dimensional flats
in $\Re^d$ and a parameter $r \geq 1$,  there exists
a nonzero $d$-variate polynomial $P$, of degree at most $r$, such
that each of the $O(r^d)$ cells induced by the zero set $Z(P)$ of $P$ (i.e.,
each component of $\Re^d \setminus Z(P)$)
intersects $O(r^{k-d} n)$ flats of $\H$. Note that for the
case $k = d-1$, this is a ``polynomial partitioning'' version
of Theorem~\ref{thm:cuttings}.

\Bnn{ROUNDING FRACTIONAL SYSTEMS}

\noindent
We now present two uses of $\eps$-nets in rounding fractional systems
to integral ones---as before, one will be algorithmic and the other combinatorial.
Given a set system $(X, \R)$, the \emph{hitting set problem} asks
for the smallest   set $Y \subseteq X$ that intersects
all sets in $\R$. Let $\opt_{\R}$ be the size of a minimum
hitting set for $\R$. Given a weight function $w: X \rightarrow \Re^+$ with $w(x) > 0$
for at least one $x \in X$, we say that $N \subseteq X$ is an $\eps$-net {\em with respect
to} $w(\cdot)$ if $N \cap R \neq \emptyset$ for any $R \in \R$ such that $w(R) \geq \eps \cdot w(X)$.
The construction of an $\eps$-net with respect to weight function $w(\cdot)$ can be reduced to the construction of a regular $\eps$-net for a different set system $(X', \R')$; the main idea is that for each
$x \in X$ we include multiple ``copies'' of $x$ in the base set $X'$, with the number of copies being proportional to $w(x)$. Using this reduction, many of the results on $\eps$-nets carry over to $\eps$-nets with respect to a weight function.

\theorem{\cmrx\cite{BG95,L01,ERS05}}
{Given $(X, \R)$, assume there is a function $f: \Re^+ \rightarrow \Ne^+$ such that for any $\eps > 0$ and weight function $w: X \rightarrow \Re^+$, an $\eps$-net of size at most $\frac{1}{\eps} \cdot f(\frac{1}{\eps})$ exists with
respect to $w(\cdot)$. Further
assume a net of this size can be computed in polynomial time. Then
one can compute a $f(\opt_{\R})$-approximation to the minimum hitting
set for $\R$ in polynomial time, where $\opt_{\R}$ is the size
of a minimum hitting set for $\R$.}
The proof proceeds as follows: to each $p \in X$ assign
a weight $w(p) \in [0,1]$ such that the total weight $W = \sum_{p \in X} w(p)$
is minimized, under the constraint that $w(R) = \sum_{p \in R} w(p) \geq 1$ for
each   $R \in \R$. Such weights can be computed in polynomial time using linear programming.
Now a $\frac{1}{W}$-net (with respect to the weight function
$w(\cdot)$) is a hitting set for $\R$; crucially,
as $W \leq \opt_{\R}$, this net is of
size at most $W f\big( W \big) \leq
\opt_{\R} \cdot f\big(\opt_{\R}\big)$.
In particular, when the set system has   $\eps$-nets
of size $O(\frac{1}{\eps})$,
one can compute a constant-factor approximation to the minimum
hitting set problem; e.g.,  for the geometric minimum hitting set problem for points
and disks in the plane. Furthermore these algorithms
can be implemented in near-linear time~\cite{AP14,BMR15}.	
When the elements of $X$ have costs, and the goal is to minimize
the cost of the hitting set, Varadarajan~\cite{V10} showed
that $\eps$-nets imply the corresponding approximation factor.

\theorem{\cmrx\cite{V10}}
{Given $(X, \R)$ with a cost function $c: X \rightarrow \Re^+$,
 assume that there
exists a function $f: \Ne \rightarrow \Ne$
such that for any $\eps > 0$ and weight function $w: X \rightarrow \Re^+$, there
is an $\eps$-net with respect to $w (\cdot)$ of cost at most $\frac{c(X)}{\eps n} \cdot f(\frac{1}{\eps})$.
Further assume such a net can be computed in polynomial time. Then
one can compute a $f(\opt_{\R})$-approximation to the minimum cost hitting
set for $\R$ in polynomial time.}

We now turn to a combinatorial use of $\eps$-nets in rounding.
A set $\C$ of $n$ convex objects in $\Re^d$ is said to satisfy
the $\HD(p,q)$ property if for any set $\C' \subseteq \C$ of
size $p$, there exists a point common to at least $q$ objects in $\C'$ (see
Chapter~\chapHelly).
Answering a long-standing open question, Alon and Kleitman~\cite{AK92}
showed that then there exists a hitting set for $\C$ whose
 size is a function
of only $p, q$ and $d$---in particular, independent of $n$. The resulting
function was improved to give the following statement.

\theorem{\cmrx\cite{AK92,KST16}\tindex{Hadwiger-Debrunner pq theorem@Hadwiger-Debrunner $(p,q)$-theorem}}
{Let $\C$ be a finite set of convex objects in $\Re^d$,
and $p,q$ be two integers, where $p \geq q \geq d+1$,  such
that for any set $\C' \subseteq \C$ of
size $p$, there exists a point in $\Re^d$ common to at least $q$ objects in $\C'$.
Then there exists a hitting set for $\C$ of size
%$O\big( (pq)^{d\frac{q-1}{q-d}}  \log^{d^3 \log d} (pq)\big)$.
$O\big( p^{d\frac{q-1}{q-d}}  \log^{c' d^3 \log d} p \big)$,
where $c'$ is an absolute constant.}

We present a sketch of the proof. Let $P$ be a point set  consisting
of a point from each cell of the arrangement of $\C$.  For each
$p \in P$, let $w(p)$ be the weight assigned to $p$ such that
the total weight $W = \sum_p w(p)$ is minimized, while satisfying
the constraint that each $C \in \C$ contains points of total weight at least $1$.
Similarly, let $w^*(C)$ be the weight
assigned to each $C \in \C$ such that the total weight
$W^*  = \sum_C w^*(C)$ is maximized, while
satisfying the constraint that each $p \in P$ lies in objects of total
weight at most $1$.
Now linear programming duality implies that $W = W^*$,
and crucially, we have $c \cdot W^* \leq 1$
for some constant $c>0$: using the $\HD(p,q)$ property,
a straightforward counting argument
shows that there exists a point $p \in P$ hitting objects in $\C$
of total weight at least $c \cdot W^*$, where $c>0$ is a constant depending
only on $p, q$ and $d$. Thus $W = W^* \leq \frac{1}{c}$, and so a weak $c$-net
for $P$ (with respect to the  weight function $w(\cdot)$) induced by convex objects
hits all objects in $\C$, and has size $O\big( \frac{1}{c^d} \log^{\Theta(d^2 \log d)} \frac{1}{c} \big)$
by Theorem~\ref{thm:weakepsnets}.
This idea was later used in proving combinatorial
bounds for a variety of geometric
problems; see~\cite{AK95,A98,AKMM02,MR16} for a few
examples.

%\pagebreak

\A{OPEN PROBLEMS}

\noindent We conclude with some open problems.

\begin{enumerate}\itemsep -1pt
\item  Show a lower bound
of $\Omega(\frac{1}{\eps} \log \frac{1}{\eps})$ on the size of any $\eps$-net for
the primal set system induced by lines in the plane.
\item  Prove a tight bound
on the size of weak $\eps$-nets for the primal set system
induced by convex objects in $\Re^d$. An achievable goal may
be to prove the existence
of weak $\eps$-nets of size $O\big( \frac{1}{\eps^{\lceil d/2 \rceil}} \big)$.
\item   Improve the current best bound of $O\big(\frac{1}{\eps} \log \log \frac{1}{\eps} \big)$ for weak $\eps$-nets for the primal set system induced
by axis-parallel rectangles in $\Re^2$.
\item   Show a lower bound of $\big( \frac{d}{2} - o(1) \big) \frac{1}{\eps} \log \frac{1}{\eps}$ for the size of any $\eps$-net for the primal set system
induced by half-spaces in $\Re^d$.
\item   Show a lower bound of $\Omega \big(\frac{1}{\eps} \log \frac{1}{\eps} \big)$ for $\eps$-nets for the primal set system induced by
balls in $\Re^3$.
\item   An unsatisfactory property of many lower
bound constructions for $\eps$-nets is that the construction of
the set system depends on the value of $\eps$---typically
the number of elements in the construction is only $\Theta(\frac{1}{\eps} \log \frac{1}{\eps})$; each element is then ``duplicated'' to derive the statement
for arbitrary values of $n$. Do  constructions exist that give a lower bound  on the $\eps$-net size
for every value of $\eps$?

%\item   Prove the following conjecture generalizing
%cuttings to $k$-dimensional flats:
%given a set $\H_1$ of $n$ $k_1$-dimensional flats,
%$\H_2$ of $m$ $k_2$-dimensional flats  in $\Re^d$,
%and an integer $r$, prove that
%there exists a partition of $\Re^d$ into
%$O(r^d)$ simplices such that
%$(i)$ each simplex intersects $O\big(\frac{n}{r^{d-k_1}}\big)$ flats of $\H_1$, and
%$(ii)$ each flat in $\H_2$ intersects $O\big(r^{k_2}\big)$ simplices.
\item Improve the slightly sub-optimal bound of
Theorem~\ref{thm:cuttingsimplices} to show the following.
Let $\S$ be a set of $n$ $(d-1)$-dimensional simplices in $\Re^d$, $d \geq 3$,
and let $m = m(\S)$ denote the number of $d$-tuples of $\S$ having a point in common.
Then for any $r \leq n$, there is a $\frac{1}{r}$-cutting of $\S$ with size at most
$\displaystyle O\big(r^{d-1} + \frac{mr^d}{n^d}\big)$.
\item Improve the current bounds for $\eps$-approximations
for the primal set system
induced by balls in $\Re^d$ to $O\big(\frac{1}{\epsilon^{2 - \frac{2}{d+1}}}\big)$.
\item Let $(X, \R)$ be a set system with
$\varphi_{\R}(m, k) = O\big( m^{d_1} k^{d-d_1}\big)$, where $1 < d_1 \leq d$ are constants (with $\varphi_{\R}(m, k)$ as defined in the first section).
Do there   exist relative $(\eps, \delta)$-approximations of
size $O\Big(   \frac{1 }{ \eps^{\frac{d+d_1}{d+1}} \delta^{\frac{2d}{d+1}} } \Big)$
for $(X, \R)$?
%Currently, the best bounds have extra logarithmic factors.
%What about of size  $O\Big(   \frac{ \log \frac{1}{\eps} }{ \eps \delta^{\frac{2d}{d+1}} } \Big)$ for $(X, \R)$ with $\pi_{\R}(m) = O(m^d)$?

\end{enumerate}

\vspace{-0.5pc}

\A{SOURCES AND RELATED MATERIALS}

\vspace{-0.5pc}

\Bnn{READING MATERIAL}

\noindent See Matou\v{s}ek~\cite{M98a} for a survey on VC-dimension, and its
relation to discrepancy, sampling and approximations of geometric set systems.
An early survey on $\eps$-nets was by Matou\v{s}ek~\cite{M93b},
and a more general one on randomized algorithms
by Clarkson~\cite{C92}.
Introductory expositions  to $\eps$-approximations and $\eps$-nets can be found in the books by Pach and Agarwal \cite{PACG}, Matou\v{s}ek~\cite{MatLectures}, and Har-Peled \cite{HarBook}. The monograph of Har-Peled \cite{HarBook} also discusses sensitive approximations and relative approximations. The books by Matou\v{s}ek~\cite{MatBook} on geometric discrepancy and by Chazelle~\cite{Cha00} on the discrepancy method give a detailed account of some of the material in this chapter. From
the point of view of learning theory,
a useful survey on approximations is Boucheron \etal~\cite{BBL05},
while the books by Devroye, Gy\"orfi, and Lugosi~\cite{DGL96}
and  Anthony and Bartlett~\cite{AB09} contain detailed proofs on random sampling
for set systems with bounded VC-dimension.
For spatial partitioning and its many applications,
we refer the reader to the book by Guth~\cite{G16}.

\Bnn{RELATED CHAPTERS}

%\noindent Chapter \phantom{1}\chapNets: Polynomial partitions

\noindent Chapter \chapDiscrepancy: Geometric discrepancy theory and uniform distribution

\noindent Chapter \chapRange: Range searching

\noindent Chapter \chapRandomization: Randomization and derandomization

\noindent Chapter \chapCoreset: Coresets and sketches

\vspace{1pc}

\Refh

\small

%\bibliography{epsapproximationsandnets}

\newcommand{\etalchar}[1]{$^{#1}$}

%\indexprologue{\noindent The name of each author cited in a chapter appears only once with a reference to that chapter; either to its first appearance in the chapter's bibliography, or, if not cited there, to its first appearance in the text of the chapter.}
%\printindex[auth]
%\printindex[term]

\end{document}